\documentstyle[12pt]{article}

       \def\de{depth}

 \let\be=\beta \let\g=\gamma \let\de=\delta
\let\e=\varepsilon   
 \let\k=\kappa \let\la=\lambda \let\m=\mu
\let\n=\nu \let\x=\xi \let\p=\pi  \let\s=\sigma
 
\let\ph=\varphi  \let\PH=\Phi 
  
 \let\G=\Gamma 

\def\0{\over } \def\1{\vec }     \def\2{{1\over2}} \def\4{{1\over4}}
\def\5{\bar }  \def\6{\partial } \def\7#1{{#1}\llap{/}}
\def\8#1{{\textstyle{#1}}}       \def\9#1{{\bf {#1}}}
 \def\llp{\hbox to 0pt{\hss /\hskip1.5pt}}
\def\llo{\hbox to 0.2pt{\hss /}} \def\llq{\hbox to 0pt{\hss /\hskip0.5pt}}
\def\so{\supset\hbox to 0pt{\hss $\displaystyle -$\hskip1pt}}

\def\<{\langle } \def\>{\rangle }

\let\nn=\nonumber
\def\bea{\begin{eqnarray}} \def\eea{\end{eqnarray}}
\def\beann{\begin{eqnarray*}} \def\eeann{\end{eqnarray*}}
\def\beq{\begin{equation}} \def\eeq{\end{equation}}

\begin{document}

%\setlength{\baselineskip}{15pt}
%\setlength{\parindent}{20pt}
%\setlength{\parskip}{6pt}
%\sloppy

\def\emph{m_{\ph}} \def\emch{m_{\chi}}
\def\mph2{m_{\ph}^2} \def\mch2{m_{\chi}^2}
\def\e2{e^2} \def\ml{m_{L}} \def\mt{m_{T}}
\def\dml2{\de m_{L}^{2}} \def\dmt2{\de m_{T}^{2}}
\def\tq{\big({e^2\0 4}+{\la\0 3}\big)\big(T^2-\Tb^2\big)}
\def\zb{Z_{\be}}
\date{}

{\large\rm DESY 93-021}\hfill{\large\tt ISSN 0418-9833}

{\large\rm February 1993}\hfill\vspace*{2cm}
\begin{center}
{\bf {\large The weak electroweak phase transition}} \\
\vspace{0.5truecm}
W. Buchm\"{u}ller, Z. Fodor\footnote{Humboldt Fellow,
on leave from Institute for Theoretical Physics,
E\"otv\"os University, Budapest, Hungary},
T. Helbig and
D. Walliser\footnote{Present address: Fermilab, Batavia, IL 60510, USA}\\
\normalsize\it Deutsches Elektronen-Synchrotron DESY, Hamburg, Germany
\vspace*{2cm}
\end{center}

\begin{abstract}
We present a detailed analysis of the phase transition
in the standard model at finite temperature.
Using an improved perturbation theory, where plasma masses are
determined from a set of one-loop gap equations,
we evaluate the effective potential $V_{eff}(\varphi,T)$ in next-to-leading
order, i.e.,
including terms cubic in the gauge coupling $g$,
the scalar self-coupling $\lambda^{1/2}$
and the top-quark Yukawa coupling $f_t$.
The gap equations yield a
non-vanishing magnetic plasma mass for the gauge bosons,
originating from the non-abelian self-interactions.
We discuss in detail size and origin of
higher order effects and conclude that the phase transition
is weakly first-order up to Higgs masses of about $70\ GeV$,
above which our calculation is no longer self-consistent.
For larger Higgs masses  even an approximation
containing all $g^4$ contributions to $V_{eff}$ is not
sufficient, at least a full calculation to order $g^6$ is needed.
These results turn out to be rather insensitive to
the top-quark mass in the range $m_t=100\ -\ 180\ GeV$.
Using Langer's theory of metastability we calculate
the nucleation rate of critical droplets and discuss some
aspects of the cosmological electroweak phase transition.
\end{abstract}
\vfill\eject

\section{Introduction}
In the standard model the electroweak gauge symmetry is spontaneously
broken. However, at sufficiently high temperatures,
above a critical temperature of about $100\ GeV$,
the  $SU(2)\times U(1)$ symmetry of weak and electromagnetic interactions
is restored \cite{wb1}-\cite{wb2}.
Only many years after the first studies of symmetry restoration
it was realized that in the standard model
the rates of anomalous baryon- and lepton-number
violating processes are unsuppressed at high temperatures \cite{wb4}.
This has important cosmological implications. In particular,
it opens the possibility to understand,
at least in principle, the generation of the
baryon asymmetry of the universe within the standard model.
Clearly, the study of "electroweak baryogenesis"
requires a detailed knowledge of the transition
from the symmetric to the broken phase in the standard model.
Recently, this has led
to a renewed interest in the electroweak phase transition
\cite{wb5}-\cite{wb9a}.

Despite the fact that several important steps
towards the understanding of the phase transition have already been made, we
are
still far from a complete description of this intriguing phenomenon.
To a large extent this is due to the infrared problems
of perturbative  finite-temperature field
theory. One manifestation of
these infrared problems is the appearance of spurious terms in
the effective potential $V_{eff}(\ph)$, which are linear in the
Higgs field $\ph$ \cite{wb8}.
On the other hand, it has been argued (cf. \cite{wb6,u1,zwirner2})
that the effective potential does not have a linear term.
A related problem is the contribution of high order loop diagrams
to low orders in the coupling constants. This can be understood
in terms of dynamically generated plasma masses which damp
infrared divergences. Much work has been devoted to the
summation of daisy, superdaisy and other types of diagrams which
yield higher order corrections to the effective potential
\cite{wb7}-\cite{parwani}. The free energy has also been evaluated
in the $1/N$-expansion \cite{jain} and for background fields
averaged over a finite volume \cite{wetterich}.
Another important question
concerns the decay of a metastable phase via nucleation, growth and
coalescence of critical droplets. Furthermore, in the case of a weak
first-order transition  subcritical
droplets and large thermal fluctuations
are of importance \cite{wb9a}.

In a recent paper \cite{u1}, the
phase transition in scalar electrodynamics was studied in detail.
Using  an improved perturbation theory, where plasma masses were
incorporated from the beginning,
the effective potential was calculated to order
$e^3$ and $\lambda^{3/2}$, and it was explicitly demonstrated
that the spurious linear terms cancel. A complete
set of one-loop gap equations was evaluated and used to determine a
region in the plane of couplings ($e^2$, $\lambda$) where the symmetric
phase is metastable. Following the theory of Langer \cite{wb16}
the nucleation rate of critical droplets was calculated,
and it was shown that a cosmological phase transition
would have been first-order up to
Higgs boson masses of the order of the vector boson mass.

In this paper we extend
the approach of ref. \cite{u1} to the
standard model. We study an $SU(2)$ gauge theory, i.e., the case $g'=0$.
This corresponds to the
approximation where the $W$-bosons and the $Z$-boson are degenerate in mass.
We take into account the three generations of fermions, with left-handed
doublets and right-handed singlets. The only relevant Yukawa coupling
is the top-quark coupling $f_t$. The effect of the
other Yukawa-couplings is negligible.

The outline of the paper is as follows. In sect. 2
we introduce our notation and study one- and two-loop contributions to the
finite-temperature effective potential. We then discuss
the breakdown of the ordinary perturbative approach and stress
the need for an improved perturbation theory. Sect. 3
is devoted to the structure of the vacuum polarisation tensor
in spontaneously broken gauge theories at finite temperature.
In sect. 4 we evaluate the effective potential.
We perform a ring summation,
thus collecting all terms of order $g^3$ and $\lambda^{3/2}$,
and we obtain a  complete set of one-loop gap equations for
the $SU(2)$ gauge theory at finite temperature,
which yield a non-vanishing magnetic plasma mass.
Using the iterative solution of these gap equations we
evaluate the effective
potential up to cubic terms in the couplings.
The equivalence of this method to the ring summation is shown,
and some properties of the potential are studied analytically.
Sect. 5 contains a detailed analysis of higher order effects.
We study the convergence of the improved perturbation theory,
the influence of the obtained transverse plasma mass
on the surface tension and the effect of superdaisy diagrams
to order $g^4$ and $\lambda^2$.
Following the theory of Langer \cite{wb16}
we compute the nucleation
rate of critical droplets in sect. 6, where we also determine the
nucleation temperature of a cosmological phase transition.
A summary of our results and concluding remarks comprise sect. 7.

Three appendices deal with matters peripheral to the main line of our
discussion. In appendix A we prove
that the ring summation with
self-energy insertions at zero momentum
yields all terms of order $g^3$ and
$\lambda^{3/2}$ in the effective potential. Appendix B gives
the result of the "superdaisy"-type summation
to order $g^4$ and $\lambda^2$. Appendix C contains some details needed
for the computation of the nucleation rate.

\section{Conventional perturbative approach}\label{breakdpt}

Let us consider the $SU(2)$ gauge theory described by the Lagrangian
${\cal L}$ depending on
a set of bosonic and fermionic fields,
\beq\label{theory}
{\cal L}={\cal L}_{gauge}+{\cal L}_{Higgs}+{\cal L}_{fermion}+
{\cal L}_{gauge \ fixing}+{\cal L}_{ghost}.
\eeq
Using standard conventions the bosonic part reads
\bea
{\cal L}_{gauge}+{\cal L}_{Higgs}
=-{1 \over 4}F^a_{\mu\nu}F^{a\mu\nu}
 +(D_\mu\Phi)^{\dag} (D^{\mu}\Phi)-\mu(\Phi^{\dag} \Phi)
-\lambda(\Phi^{\dag} \Phi)^2, \nn \\
\mu<0,\quad D_{\mu}=\partial_\mu-ig{\tau^a \over 2}W_\mu^a,
\quad a=1,2,3, \qquad
\eea
where the Higgs field $\Phi$ is an $SU(2)$ doublet.
The fermionic part is given by
\bea
{\cal L}_{fermion}
=\bar \psi_Li\gamma^{\mu} D_\mu \psi_L
 +f_t{\overline {\left( \begin{array}{c} t \\ b \\  \end{array} \right)} _L}
(i\tau_2\Phi^*)t_R+h.c.
\eea
Here $\tau^a$, $a=1,2,3$, denote the Pauli matrices, $\psi_L$ stands for all
left-handed fermions and $f_t$ is the Yukawa coupling of the top-quark.
All other Yukawa couplings are much smaller and can be neglected.
The gauge-fixing and ghost Lagrangians read
\bea
{\cal L}_{gauge\ fixing} &=& -{1 \over 2\eta}G_aG^a\,, \quad
G_a=\partial_\mu W^\mu_a-
{1 \over 2}\eta g \ph \chi_a\,, \\
{\cal L}_{ghost} &=&\bar c_aM^{ab}c_b\,, \qquad
\delta G^a=M^{ab}\delta \omega_b\,;
\eea
here the $SU(2)$ doublet $\Phi$ of scalar fields has been written as
\beq
\Phi={1 \over \sqrt{2}}
\left(
\begin{array}{c}
\chi_1+i\chi_2 \\
\ph+h+i\chi_3 \\
\end{array}
\right) \quad,
\eeq
where $h$ is the Higgs field, $\chi_a \ (a=1,2,3)$ are the three Goldstone
bosons and $\ph$ is a constant background field.
In the following we will use Landau gauge, i.e., we take the limit
$\eta \rightarrow 0$.
The classical minimum is at $\ph=v\equiv\sqrt{-\mu/\la}$.
The tree-level vector boson mass $m$, the top-quark mass $m_t$,
the Higgs boson mass $\bar m_\ph$ and the Goldstone boson mass
$\bar m_\chi$ are given by
\bea
m^{2}={g^2\ph^2 \over 4},\qquad
m_{t}^2={f_t^2 \over 2}\ph^2, \qquad \\
\bar{m}_\ph^2=\la(3\ph^2-v^2),
\qquad\bar{m}_\chi^2=\la(\ph^{2}-v^{2}).
\eea
Note that $\bar{m}_\ph^2$ and $\bar{m}_\chi^2$ are negative for small values
of $\ph$. The connection between couplings and zero-temperature masses reads
$g=2m_W /v$, $f_t=\sqrt{2}m_t /v$ and $\la=m_H^2/2v^2$, where
$v=246 \ GeV$.

The finite temperature action is given by the integral
\beq
S_\beta=\int_\beta dx{\cal L},
\eeq
with
\beq
\int_\beta dx\equiv\int_0^\beta d\tau\int d^3x,\ \ \ \beta={1 \over T}.
\eeq
Boson and fermion fields have to satisfy
periodic and antiperiodic boundary conditions in $\tau$, respectively.

The finite-temperature effective potential can be perturbatively
evaluated in the loop expansion (cf. ref. \cite{wb12}).
The one-loop contribution is shown in fig. \ref{oneloop}.
Its temperature dependent part can be calculated using standard
techniques. The sum of the leading term in the high temperature
expansion and the tree-level potential yields:
\bea \label{pot1loop0}
V_0(\ph)+V_1^{(1)}(\ph,T)=
{1 \over 2}\left({3g^2+8\lambda+4f_t \over 16}T^2
-\lambda v^2 \right) \ph^2+{1 \over 4}\lambda \ph^4.
\eea
As is well known, this potential predicts a second-order phase transition.
Note, that the finite-temperature contribution due to scalar
loops is proportional to  $\lambda$ and therefore of the same order as the
tree-level term $\lambda \ph^4/4$.

Going one order further in the high-temperature expansion one obtains
\bea \label{pot1loop}
V_0(\ph)+V_1^{(2)}(\ph,T)=
&&{1 \over 2}\left({3g^2+8\lambda+4f_t \over 16}T^2
-\lambda v^2 \right) \ph^2+{1 \over 4}\lambda \ph^4    \nn   \\
&&-\left(9 m^3+\bar m_\ph^3 +3\bar m_\chi^3 \right){T\over 12\pi}\,.
\eea
The potentials (\ref{pot1loop0}) and (\ref{pot1loop})
differ in several important respects:\hfill\break
(i) We are faced with the familiar problem that
$\bar m_\ph$ and $\bar m_\chi$ become imaginary for small
values of $\ph$, i.e., the
naive one-loop effective potential is complex. This phenomenon is not an
artefact of the high-temperature expansion, since the
exact one-loop contribution
of the Higgs field is proportional to
\beq \label{highT}
J_+(m_\ph^2)=\int_0^\infty dx x^2 log\left(1-
e^{-\sqrt{x^2+(3 \ph^2-v^2)\lambda/T^2}}\right),
\eeq
which is also complex for small values of $\ph$. A
similar expression is obtained for the Goldstone fields,
and no cancellation appears between the two contributions.
At zero temperature, the imaginary part of the potential has been
related to the lifetime of a particular quantum state
\cite{ewein}.\hfill\break
(ii) The new cubic  terms are of order $g^3$ and $\lambda^{3/2}$,
and thus of
higher order than the finite-temperature corrections in $V^{(1)}_1$.
Clearly, these new contributions must be combined with terms of the
same order which appear in higher order loop diagrams. \hfill\break
(iii) The cubic terms originate from integrals which are infrared
divergent for vanishing masses. No cubic term results
from the top-quark loop, since the fermionic modes always have non-zero
Matsubara frequencies ($\omega_n=(2n+1)\pi T$), and hence do not suffer
from infrared problems. \hfill\break
(iv) In order to see the qualitative features of the
effective potential (\ref{pot1loop})
we first ignore the terms proportional to $\lambda^{3/2}$,
which can be justified for $\lambda\ll g^2$.
For temperatures above a certain "barrier" temperature $T_b$,
the second derivative of this modified potential
is positive, thus
the symmetric phase is a local minimum of the effective potential.
Here the barrier temperature $T_b$ is
\beq\label{TB}
T_b^2={16\lambda v^2 \over 3g^2+8\lambda +4f_t^2}\,.
\eeq
In eq. (\ref{pot1loop}) the terms proportional to $\ph^3$
and $\ph^4$ have opposite signs.
As a result of a compensation between these two terms, for temperatures
close enough to $T_b$, another minimum exists which, at a critical
temperature $T_c$, is degenerate with the minimum at $\ph=0$. Hence, in this
approximation the phase transition is first-order.

According to (ii) we have to consider higher
loop terms involving boson fields, first the two-loop contributions
shown in fig. \ref{twoloop}. These graphs yield linear terms in the
temperature dependent part of the effective potential:
\bea
V_2^{(scalar)}(\ph,T) =&&
- {1 \over 128\pi}(3 g^2 + 8 \lambda + 4 f_t^2)
(\bar m_\ph+3\bar m_\chi)T^3 \nn \\
&&+ {\cal O}(g^4,\lambda^2) \quad,  \\
V_2^{(vector)}(\ph,T) =&&
- { 11 \over 32\pi } g^3 \ph T^3 \ +\ {\cal O}(g^4,\lambda^2) \quad.
\eea
The expressions are cubic in the couplings,
and therefore they must be combined with the
contributions of the same order in $V^{(2)}_1(\ph,T)$.
However, the combined
result does not solve any of our previous problems related to the scalar
sector, but rather introduces a new problem,
the appearance of a very disturbing linear term
in the effective potential. The fact that the two-loop
diagrams yield terms of order $g^3$ and $\lambda^{3/2}$,
and not only $g^4$ and $\lambda^2$ is another manifestation
of the infrared problems in finite-temperature field theories. The
two-loop contributions can be considered as one-loop graphs where the other
loop plays the role of a self-energy insertion.
The infrared divergencies are cut off by these
dynamically generated masses (proportional to $g$ or $\sqrt \lambda$),
which leads to factors $T/m_i$ and thereby reduces
the order in $g$ or $\lambda$ of the corresponding diagram.

In appendix A we show that to order $g^3$ and $\lambda^{3/2}$
only ring diagrams contribute (cf. fig. \ref{ring}).
The complete
summation of these terms with the proper combinatoric
factors is required to obtain the correct effective potential to order
$g^3$ and $\lambda^{3/2}$. Therefore, we need the self-energy terms
at zero external momentum. The evaluation of these contributions
is straightforward for scalar fields, but for vector fields
the polarisation tensor needs special
attention in spontaneously broken gauge theories. The next section is
addressed to this question.

\section{The structure of the gauge boson propagator}\label{improvedpt}

Most of the results of this section have already been discussed
in ref. \cite{u1}. Nevertheless, they are included here to make our
paper self-contained.

In order to determine the plasma masses of vector and scalar fields, we first
have to discuss the structure of the vector propagator at finite temperature.
The gauge boson self-energy $\Pi_{\m\n}(k)$ depends on the 4-momentum
$k^{\m}$ and the 4-vector $u^{\m}=(1,\vec{0})$ which specifies the rest frame
of the system (cf. ref. \cite{wb12}). Hence, in general $\Pi_{\m\n}$ is a
linear combination of four tensors. A convenient choice is
\bea
P_{T\,\m\n}&=&
{{ g}_{\m}}^{i}\left(\de_{ij}-\frac{k_{i}k_{j}}{\vec{k}^{2}}\right)
{{ g}^{j}}_{\n},\\
P_{L\,\m\n}&=&
\frac{k_{\m}k_{\n}}{k^{2}}-{ g}_{\m\n}-P_{T\,\m\n}
\ = \ \frac{k^{2}}{\vec{k}^{2}}u^{T}_{\m}u^{T}_{\n},\\
P_{G\,\m\n}&=&
-\frac{k_{\m}k_{\n}}{k^{2}},\\
S_{\m\n}&=&{\frac{1}{\sqrt{2\vec{k}^2}}}
\left(k_{\m}u^T_{\n}+k_{\n}u^T_{\m}\right),
\eea
where $u^{T}_{\m}=u_{\m}-k_{\m}\frac{u\cdot k}{k^{2}}$ is transverse,
$u^T_\m k^\m=0$. These tensors satisfy
the relations
\bea
&&P_{T}^{2}\,=\,-P_{T},\quad P_{L}^{2}\,=\,-P_{L},\quad
P_{G}^{2}\,=\,-P_{G},\quad S^{2}\,=\,\frac 1 2(P_{L}+P_{G}),\label{p5}\\
&&P_{T}P_{L}\,=\,P_{T}P_{G}\,=\,P_{L}P_{G}\,=\,SP_{T}\,=\,
P_{L}SP_{L}\,=\,0,\label{p6}\\
&&{P_{T\,\m}}^{\m}=2{P_{L\,\m}}^{\m}=2{P_{G\,\m}}^{\m}=-2,\quad
{S_{\m}}^{\m} = {P_{L\,\m}}^{\n}{S_{\n}}^{\m} = 0 \ .\label{p7}
\eea
In Landau gauge, where the bare propagator is transverse,
\beq
D(k)=\frac{-1}{k^2-m^2}(P_L+P_T),
\eeq
the gauge-boson self-energy tensor
\beq\label{vacpolten}
\Pi(k)=\Pi_{L}(k)P_{L}+\Pi_{T}(k)P_{T}+\Pi_{S}(k)S+\Pi_{G}(k)P_{G}
\eeq
yields the full propagator
\bea\label{fullprop}
\tilde{D}(k)&=&\sum_{n=0}^{\infty}\,D(k)\big[\Pi(k)D(k)\big]^{n}\nn\\
&=&\frac{-1}{k^2-m^2-\Pi_L(k)}P_L+\frac{-1}{k^2-m^2-\Pi_{T}(k)}P_{T}.
\eea
Due to the relations (\ref{p5}) and (\ref{p6}) the full propagator
does not depend on $\Pi_S(k)$ and $\Pi_G(k)$.

However, knowledge of $\Pi_G$ is important since it enters in the
relations which yield the longitudinal and transverse plasma masses:
\bea
\dml2 &=&\Pi_L(0)\ =\ \mbox{Tr}[\Pi(0)P_L]\ =\ -\Pi_{00}(0),\label{plasmaml}\\
\dmt2 &=&\Pi_T(0)\ =\ \2\mbox{Tr}[\Pi(0)P_T] \nn \\
 &=&\ -\frac{1}{2}\left[{\Pi^{\m}}_{\m}(0)+\Pi_L(0)
+\Pi_G(0)\right]\ ,\label{plasmamt}
\eea
where
\beq\label{defpi}
\Pi_G(k)=\mbox{Tr}[\Pi(k)P_G].
\eeq
Here $\Pi_{\m\n}(0)$ is defined by setting first $k_{0}=0$
and then performing the
limit $\vec{k}^{2}\rightarrow 0$. In gauge theories with unbroken symmetry one
has $\Pi_G=0$ in Landau gauge.
Note, that this is \underline{not} the case if
the symmetry is spontaneously broken. As eq. (\ref{plasmamt}) shows this fact,
which seems to have gone unnoticed in the literature,
is important in order to extract the correct transverse mass
from $\Pi_{\m\n}$.
As we will see, in non-abelian gauge theories
this transverse mass turns out to contain a field independent
magnetic plasma mass which plays an important role in the
estimation of the higher order effects. Hence, the unambiguous determination
of this term is needed.

\section{The effective potential in next-to-leading  order}

\centerline {\bf A. Ring summation}

\bigskip

The one-loop gauge boson self-energy corrections
for the $SU(2)$ gauge theory are
shown in fig. (\ref{selfenergy}).
The self-energy contributions have very complicated dependence on the momentum
($k_0,|\vec{k}|$). As we show in appendix A,
to order $g^3$ and
$\lambda^{3/2}$ one has to set $k_0=0$ for the external
lines of the self-energy diagrams in the ring summation. On the other
hand, by use of a Taylor expansion in $|\vec{k}|$ it is easy
to show that only the $|\vec{k}| \rightarrow 0$ limit
of the self-energy graphs contribute to the effective potential
in this order.
In this limit we find for the longitudinal and transverse
plasma masses to order $g^3$ and $\lambda^{3/2}$:
\bea
\delta m_L^2&=&{11 \over 6}g^2T^2-{g^2 \over 16\p}\left({4m^2\over
m+\bar{m}_{\ph}}
+\bar{m}_{\ph} +3\bar{m}_{\chi}+16m\right)T,
\label{dmul1}\\
\delta m_T^2&=&{g^2T\over 3\pi}m-{g^2\over 6\p}
\left({m^2\over m+{\bar m}_{\ph}}
-{1 \over 8}\frac{({\bar m}_{\ph}-{\bar m}_{\chi})^2}
{{\bar m}_{\ph}+{\bar m}_{\chi}}\right)T.
\label{dmut1}
\eea
Here we have neglected terms of higher order in
the high-temperature expansion,
since we will not need the corresponding terms in the effective potential.
It is worth  mentioning that no field independent
transverse plasma mass appears in eq. (\ref{dmut1}),
in accord with the well-known
fact that in one-loop order it vanishes in all gauges. Having also determined
the corresponding scalar self-energy contributions, one can perform
the ring summation. The corresponding diagrams are shown in fig. \ref{ring}
where the "blobs" stand for one-loop self-energy contributions (cf.
fig. \ref{selfenergy} and similar scalar terms). Special care
is needed with respect to the infrared limit
and the combinatoric factors, particularly in the case of
one self-energy insertion, since these graphs are two-loop
diagrams with different symmetries depending on the type of
propagators. The ring summation yields the potential
\bea \label{thepotential}
V_{ring}(\ph,T)&=&{1 \over 2}\left({3g^2\over 16}+{\la\0 2}
+{1 \over 4}f_t^2\right)
(T^2-T_b^2)\ph^2+{\la\over 4}\ph^4  \nn\\
        & &-(3m_L^3+6m_T^3+m_\ph^3+3m_\chi^3)
\frac{T}{12\p}+{\cal O}(g^4,\la^2,f_t^4),
\eea
where $T_b$ is given by eq. (\ref{TB}) and
the masses are the sum of tree-level terms and one-loop self-energy
corrections to order $g^2$ and $\lambda$
\bea
m_L^2&=&{11\over 6}g^2 T^2+{g^2\ph^2 \over 4},\nn\\
m_T^2&=&{g^2\ph^2 \over 4},\nn\\
m_\ph^2&=&\left({3\over 16}g^2+{\la\over 2}+{1 \over 4}f_t^2\right)
(T^2-T_b^2)+3\lambda\ph^2,\nn\\
m_\chi^2&=&\left({3\over 16}g^2+{\la\over 2}+{1 \over 4}f_t^2\right)
(T^2-T_b^2)+\lambda\ph^2.
\eea

The potential (\ref{thepotential}) describes a first-order phase transition,
since there exist two degenerate minima at a critical temperature $T_c$,
similar to the case considered in sect. 2. However,
due to the non-vanishing longitudinal plasma mass the strength of the
transition is less than the high-temperature expansion
of the one-loop result suggests. Clearly, a non-zero
field independent magnetic mass $m_T$ would further weaken the transition.

The potential (\ref{thepotential}) contains no term linear in $\ph$.
Hence, the linear terms which appear in the
two-loop expression are cancelled by contributions from the
ring summation. As we show in appendix A the result (\ref{thepotential})
contains all
terms of order $g^3$ and $\lambda^{3/2}$. Therefore the
appearance of linear terms
of order $g^3$ or $\lambda^{3/2}$ can be ruled out. To higher
order in $g$ and $\lambda$ the absence of a linear term has not
yet been proven. Note, that for temperatures above the barrier temperature
$T_b$
all masses, and therefore also the potential, are real.

Formal arguments concerning
the existence or nonexistence of  linear terms
in the framework of the perturbative expansion in the coupling constant
are questionable, since such an expansion does not
necessarily reflect the behaviour of the potential at the origin.
As an illustration consider a hypothetical potential
containing a term proportional to
\bea
 \sqrt{g^{2n+2}T^2+g^{2n}\ph^2}.
\eea
This function is symmetric in $\ph$ and contains no linear term.
Nevertheless, a formal expansion in powers of $g$ gives
rise to a linear term,
\bea
\sqrt{g^{2n+2}T^2+g^{2n}\ph^2} \approx
g^n\ph+{g^{n+2}T^2 \over 2\ph} + {\cal O}(g^{n+4}).
\eea
On the other hand, a potential with a term
\bea
 \ph \sqrt{g^{2n+2}T^2+g^{2n}\ph^2}
\eea
does contain a linear term which, however, is not visible
in the expansion in powers of $g$:
\bea
 \ph \sqrt{g^{2n+2}T^2+g^{2n}\ph^2} \approx g^n\ph^2+g^{n+2}T^2/2
+{\cal O}(g^{n+4}).
\eea
The behaviour of the potential obtained by an
expansion in the coupling constant does not coincide with that of the
exact potential.
Note, however, that in both cases the ratio of the
next-to-leading term and the leading term
becomes very large at $\ph \approx 0$
which indicates that the perturbative expansion
in $g$ is not valid in this region. Because of this fact we will
carefully study
the effects of the higher order terms
in the couplings at
$\ph \approx 0$ and close to the second non-trivial minimum of the effective
potential (\ref{thepotential}).

The ring summation yields a sensible result to order $g^3$ and $\lambda^{3/2}$.
This is rather surprising since individual terms in this sum are not
well defined for small values of $\ph$ due to the imaginary masses
which appear in the scalar field propagators. This inconsistency would
show up in the next higher order, i.e., $g^4$ and $\lambda^2$,
where all diagrams have to be added to the ring diagrams
which contribute to this order.

\bigskip

\centerline {\bf B. Gap equations}

\bigskip

In this subsection we shall evaluate the potential (\ref{thepotential})
in a more elegant way
which will also enable us to estimate higher order corrections to our result.
For small values of $\ph$ the temperature dependent plasma mass corrections
$\delta m^2_{L,T}$ can become larger than the tree-level mass $m^2$. This
suggests an improved perturbation theory \cite{wb2,frele},
where loop diagrams are evaluated
with boson propagators containing the exact masses
$m^{2}_{L,T}=m^{2}+\de m^{2}_{L,T}$,
$m^{2}_{\ph,\chi}=\bar{m}_{\ph,\chi}^2+\de m_{\ph,\chi}^2$. The
radiative corrections are treated as counter terms
\bea
\de S_{\be}&=&-{1\over 2}\int_{\be}dp\ \Big[\tilde{A}^{\m}(p)(\dml2
P_{L\,\m\n}+\dmt2 P_{T\,\m\n})\tilde{A}^{\m}(p)\nn\\
&&\hphantom{-{1\over 2}\int_{\be}dp\, \Big[}-\de\mph2\,\tilde{\ph}^2(p)
-\de\mch2\,\tilde{\chi}^2(p)\Big],
\eea
and are determined self-consistently by solving gap equations at the
corresponding loop order.  As it was pointed out above,
there is no need to use this procedure for fermions, since fermions
do not suffer from infrared problems.
The ghost propagator remains massless in Landau gauge. Therefore,
in our approach the bare fermion and ghost propagators play the role of full
propagators.
The one-loop contributions to the effective potential with these dressed
propagators are shown in fig. \ref{oneloopfull}.
Note, that here and in the following
the blobs stand for exact masses and not only for  one-loop
self-energy insertions.
The mass counter terms for top-quark and
ghost are zero.

The relevant one-loop graphs for the gap equations are shown
in fig. \ref{gap}.
A rather lengthy, but
straightforward calculation yields the following set of equations
to order $g^3$ and $\lambda^{3/2}$:
\bea
\ml^2&=&{11 \over 6}g^2T^2+m^2-{g^2 \over 16\p}\left({4m^2\over \ml+\emph}
+\emph +3\emch+16\mt\right)T,\label{gapeqml}\\
\mt^2&=&{g^2T\over 3\pi}m_T+m^2 -{g^2\over 6\p}
\left({m^2\over \mt+\emph}
-{1 \over 8}\frac{(\emph-\emch)^2}{\emph+\emch}\right)T,\label{gapeqmt}\\
\mph2 &=& \left({3g^2\over 16}+{\la\over 2}+{1 \over 4}f_t^2\right)
\left(T^2-T_b^2\right)+3\bar{m}^2\nn\\
&&-{3g^2\over 16\p}\left[\ml+2\mt+m^2({1\over \ml}
+{2\over \mt})\right]T\nn\\
&&-{3\la\over 4\p}\left[\emph+\emch+\bar{m}^2({3\over \emph}
+{1\over \emch})\right]T,\label{gapeqmp}\\
\mch2 &=& \left({3g^2\over 16}+{\la\over 2}+{1 \over 4}f_t^2\right)
\left(T^2-T_b^2\right)+\bar{m}^2\nn\\
&&-{3g^2\over 16\p}\left(\ml+2\mt\right)T\nn\\
&&-{\la\over 4\p}\left(\emph+5\emch
+\frac{4\bar{m}^2}{\emph+\emch}\right)T,\label{gapeqmx}
\eea
where
\beq
m\,=\,g\ph/2,\quad \bar{m}\,=\,\sqrt{\la}\ph,\quad T_b^2\,=\,
\frac{16\la v^2}{3g^2+8\la+4f_t^2}\,.
\eeq
As already mentioned,
the ghost mass remains zero, and the mass corrections for fermions
are not important since their Matsubara frequencies are always at least
${\cal O}(1)\cdot T$.
Thus, there is no need to consider
gap equations for fermions and ghosts.

It is instructive, first to consider the gap equations
for the pure scalar theory ($g=0$, $f_t=0$)
at $\ph=0$. From eqs. (\ref{gapeqmp}) and (\ref{gapeqmx}) one obtains
for temperatures close to the barrier temperature
\beq
\emph\sim\emch\sim(T-T_b),\label{dolan}
\eeq
i.e., $\emph$ and $\emch$ approach zero with critical index one. This
well-known
result was first obtained by Dolan and Jackiw in the large-$N$ limit
\cite{wb2}.
We obtain the same result, because at $\ph=0$ only graphs $ e)$
and $ l)$ of fig. \ref{gap} contribute, which are the leading terms in the
$1/N$-expansion.

Of particular interest is eq. (\ref{gapeqmt})
for the magnetic mass. At $\ph=0$, one has $m_{\ph}=m_{\chi}$,
and therefore
\beq
m_T^2={g^2T \over 3\pi}m_T.
\eeq
This equation has two solutions,
\beq m_T=0 , \eeq
and
\beq m_T={g^2T \over 3\pi},\label{mt}
\eeq
thus an unwanted ambiguity seems to appear.
However, only the second solution is physical, since only
$m_T=g^2T/(3\pi)$ can be continuously connected to a positive solution
of eq. (\ref{gapeqmt}) at $\ph>0$.
The $\ph\rightarrow 0$
limit of the negative solution of eq. (\ref{gapeqmt}) is $m_T=0$.
This negative solution is unphysical, since $m_T$ is defined as
absolute value of a square root
appearing in the high temperature expansion.
The result (\ref{mt}) has been independently derived in ref. \cite{zwirner2}.

The obtained value $m_T=g^2T/(3\pi)$ for the transverse plasma mass
goes beyond the accuracy of our calculation which is only valid
to order $g$. However, the appearance of a magnetic mass of order $g^2$
is expected to be a nonperturbative feature
of the theory \cite{linde}. In order
to study the dependence of our results on the unknown value of the
magnetic mass we will take $m_T=\g g^2T/(3\pi)$
as the $\ph\rightarrow 0$ limit.

It is well known that in the abelian Higgs model the transverse
plasma mass is zero at finite temperature, which is in agreement
with our previous results \cite{u1}. In nonabelian gauge theories the
situation is different. In this case the
gauge boson self-couplings (cf. figs. \ref{gap}q, \ref{gap}u) yield
a non-vanishing transverse plasma mass.
This magnetic mass can be self-consistently determined
by summing an infinite set of diagrams. The non-zero
solution (\ref{mt}) for the transverse plasma mass can be viewed as a result
of an infinite iteration process with an arbitrarily small initial
value. In a diagrammatic picture the iterative solution corresponds
to an infinite sum over "superdaisy"-type self-energy insertions.
The appearance of a magnetic mass term has been studied by other
methods in finite-temperature QCD \cite{kalas}.
Lattice $SU(2)$ models and infinite summations also give values
for $\gamma$ which are ${\cal O} (1)$.

Let us now solve the gap equations, which represent
a set of nonlinear equations with four variables:
\beq
{\cal M}^2= f({\cal M})\,, \quad
{\cal M}=(m_L,m_T,m_\chi,m_\ph)\,, \quad
{\cal M}^2=(m_L^2,m_T^2,m_{\chi}^2,m_{\ph}^2)\,.
\eeq
Except for linear problems, finding of roots
invariably proceeds by iteration. Starting from some approximate solution,
a useful algorithm will improve its accuracy. We are interested in
a Taylor-expansion in $g$ and $\sqrt \lambda$ of the result.
Setting $g=\sqrt\lambda=0$ gives zero for ${\cal M}$, thus the
Taylor expansion for ${\cal M}$
starts with $g$ or $\sqrt\lambda$, and for
${\cal M}^2$ with $g^2$ or $\lambda$. Suppose
that we know the results for ${\cal M}$ to order $n$. Inserting this
trial solution in the right-hand-side of the gap equations we obtain
an expression to order $n+2$ for ${\cal M}^2$,
and to order $n+1$ for ${\cal M}$. The coefficients of the
previous orders do not change. Consequently, the above iterative procedure
reproduces the Taylor-expansion of the exact solution order by order.

This iterative solution has a diagrammatical picture. The lowest order
result corresponds to the one-loop self-energy insertion at zero
external momentum. A new iteration means one order higher
in the coupling, e.g., the sum of the one-loop self-energy insertions
of the one-loop self-energy insertion, the so-called "daisy" graphs,
the next order
gives a new self-energy insertion on the bare lines,
and so on and so forth.
After an infinite number of iterations one obtains the sum of all
"superdaisy" diagrams \cite{wb2}.
Note, however, that this summation does not give
the correct answer for the effective potential to order $g^4,\ \lambda^2$
and higher. One cannot justify the zero external momentum limit of the
gap equations for higher order contributions (cf. appendix A).

Our goal is to calculate the effective potential to order $g^3$ and
$\lambda^{3/2}$. Due to the global
$SU(2)$ symmetry of the theory the potential is
only a function of
$\sqrt{2\PH^{\dagger}\PH}=\sqrt{\ph^2+\chi^2_1+\chi^2_2+\chi^2_3}$.
Hence,
at $\chi=0$, the masses $\emph(\ph,T)$ and $\emch(\ph,T)$ are given by
\bea \label{exactmass}
\mph2(\ph,T)=\frac{\6^2V(\ph,T)}{\6\ph^2}\,,\label{turn}\\
\mch2(\ph,T)={1\over \ph}\frac{\6V(\ph,T)}{\6\ph}\,\label{extrem}.
\eea
Thus, in order to obtain the potential to order $g^3$ and
$\lambda^{3/2}$ one has to evaluate $m_\chi$ or $m_\ph$ to the same order.
This means two successive self-energy insertions, which is exactly
the ring summation. The result reads:
\bea\label{thepotential1}
V_{gap}(\ph,T)&=&\int^{\ph}d\ph'\,\ph'\,\mch2(\ph',T)\nn\\
        &=&{1 \over 2}\left({3g^2\0 16}+{\la\0 2}+{1 \over 4}f_t^2\right)
(T^2-T_b^2)\ph^2+{\la\0 4}\ph^4\nn\\
& &-\left({3\ml^{(1)}}^3+6{\mt^{(1)}}^3
+{m^{(1)}_{\ph}}^3+3{m^{(1)}_{\chi}}^3\right)
\frac{T}{12\p} \nn\\
& &+{\cal O}(g^4,\la^2,f_t^4)\,,
\eea
where the lowest order masses are
\bea\label{mass0}
{\ml^{(1)}}^2&=&{{11\over 6}g^2 T^2+m^2},\nn\\
{\mt^{(1)}}^2&=&{\g^2{1 \over 9\pi^2}g^4T^2+m^2},\nn\\
{m^{(1)}_{\ph}}^2&=&{({3\over 16}g^2+{\la\over 2}+{1 \over 4}f_t^2)
(T^2-T_b^2)+3\bar{m}^2},\nn\\
{m^{(1)}_{\chi}}^2&=&{({3\over 16}g^2+{\la\over 2}+{1 \over 4}f_t^2)
(T^2-T_b^2)+\bar{m}^2}.
\eea
As it should be, this result is identical with
the ring potential (\ref{thepotential}) for $\gamma=0$.
In order to illustrate the dependence of our results on the unknown value
of the magnetic mass we have modified the iterative solution of the gap
equations to leading order in the couplings in such a way that
$m_T^{(1)}(\ph=0)=\gamma g^2T/(3\pi)$.

The evaluation of the effective potential to order $g^4$,
$\la^2$ and $f_t^4$ requires
the incorporation of two-loop contributions
in eqs. (\ref{gapeqml}) - (\ref{gapeqmx}), and
in order to obtain the exact effective potential one has to solve
the full Dyson-Schwinger equations, a non-trivial task!
However, the obtained  gap equations already contain
some part of the higher order terms, namely the "superdaisy" diagrams,
and thus contain valuable information about higher order
corrections. Before we investigate these questions in the next
section, we study some properties of the obtained potential
(\ref{thepotential1}).

\bigskip

\centerline {\bf C. Plasma masses and the order of the phase transition}

\bigskip

A nice feature of the one-loop effective potential
considered in sect. 2, with or without cubic term,
was the possibility to study the phase structure analytically.
Due to the plasma masses our potential
(\ref{thepotential}) or (\ref{thepotential1}) now
contains terms of the form $(C^2+\ph^2)^{3/2}$, and is therefore
more complicated.
Nevertheless, it is still possible to study its main
properties analytically.

In order to understand the effect of plasma masses consider a hypothetical
potential with one plasma mass term proportional to $T$,
\beq
V_{hyp}(T,\ph)={a \over 2}(T^2-T_b^2)\ph^2-{bT \over 3}(c^2T^2+\ph^2)^{3/2}
+{\lambda \over 4}\ph^4,
\eeq
where $a,b,c$ and $\lambda$
are positive real numbers, independent of temperature. $V_{hyp}$ has a local
minimum at the origin as long as its second derivative there is positive.
This condition yields a corrected barrier temperature $\tilde{T}_b$,
\beq
\tilde{T}_b^2={T_b^2 \over 1-bc/a}\,.
\eeq

The phase transition described by this potential is first-order if at
the barrier temperature $\tilde{T}_b$ a second minimum exists at non-zero
$\ph$,
otherwise it is second-order. The condition that at $\tilde{T}_b$ the first
derivative of the potential with respect to $\ph$ vanishes at
some value $\ph>0$ implies for the couplings
\beq \label{cond}
\lambda<{b \over 2c}\,.
\eeq
In the standard model only the parameter $a$ depends on the top-quark
Yukawa coupling, hence this condition does not depend
on the top mass. Clearly, it implies that
at a critical value of the Higgs mass the transition
changes from first-order to second-order.
It is easy to show that for several plasma mass terms of the form
\beq
 {b_1T \over 3}(c_1^2T^2+\ph^2)^{3/2}+
 {b_2T \over 3}(c_2^2T^2+\ph^2)^{3/2}+...
+{b_nT \over 3}(c_n^2T^2+\ph^2)^{3/2}
\eeq
the condition (\ref{cond}) for a first-order phase transition becomes
\beq\label{condition}
\lambda<{b_1 \over 2c_1}+{b_2 \over 2c_2}+...+{b_n \over 2c_n}\,.
\eeq

If one neglects in the electroweak potential (\ref{thepotential})
terms of order $\la^{3/2}$,
thus assuming $g^2\gg\lambda$, the cubic term with
longitudinal plasma mass is sufficient
to give a first-order transition if $m_H<13 \ GeV$.
However, because of the second cubic term with zero plasma mass
the phase transition is first-order for all Higgs masses.
This situation changes if a transverse
plasma mass is included, as in eq. (\ref{thepotential1}).
The corresponding term in the effective
potential yields a second-order transition if
\beq
m_H>{85 \ GeV \over \sqrt{\gamma}}\,.
\eeq
Inclusion of the cubic term with longitudinal plasma mass
slightly relaxes this bound according to condition (\ref{condition}).
As already mentioned, $\g = O(1)$.

The scalar part of the electroweak potential needs a different treatment.
Set $g=0$. Now the potential is of the form
\bea
\tilde{V}_{hyp}(T,\ph)&=&
{a \over 2}(T^2-T_b^2)\ph^2 +{\lambda \over 4}\ph^4
-{b \over 9} \Big(3(a(T^2-T_b^2)+\lambda\ph^2)^{3/2}\nn\\
&&+(a(T^2-T_b^2)+3\lambda\ph^2)^{3/2}\Big)T,
\eea
and the temperature, where the potential barrier vanishes, is
\beq
\tilde{T}_b^2={T_b^2 \over 1-4b^2\lambda^2/a}.
\eeq
The only solutions of the equation
$\partial\tilde{V}_{hyp}(\tilde{T}_b,\ph)/\partial\ph=0$ are
\beq
\ph^2=0, \ \ \ \ph^2=b\lambda \tilde{T}_b^2(-4\pm i),
\eeq
i.e., the scalar part of our potential always yields a second-order phase
transition, independent of the value of $\lambda$.

In the general case of non-zero plasma masses for vector and scalar
bosons (cf. eq. (\ref{thepotential1})),
the order of the phase transition changes at a critical Higgs mass which can
be evaluated as function of $\gamma$. The result is shown in fig.
(\ref{gamma}).
Note, that for the present lower experimental bound on the Higgs mass the
phase transition is always second-order for $\gamma>2.5$.

As we shall discuss in detail in the next section, also the improved
perturbative approach becomes inapplicable close to the barrier temperature.
Therefore, we cannot calculate the effective potential at this temperature.
Nevertheless, many features of our numerical analysis will be very
similar to the result of our analytical study in this section, which
one may expect based on continuity arguments.

\section{Higher order effects}\label{highorder}

In this section we will estimate the size of different higher order
corrections to the effective potential (\ref{thepotential1}). First we will
use an extended version of the method discussed in ref. \cite{u1}
to check the convergence of the perturbation series.
We will then study effects of a non-vanishing magnetic mass,
in particular its influence on the surface tension, and
finally we will perform an analysis including some higher order terms
in the effective potential.

\bigskip

\centerline {\bf A. Convergence of the perturbation series}

\bigskip

At the origin,
$\ph=0$, the curvature of the potential $\mph2(0,T)$ reads
(cf. eq.(\ref{gapeqmp}))
\bea
\mph2&=&({3 \over 16}g^2+{\lambda \over 2}+{1 \over 4}f_t^2)(T^2-T_b^2)
 - {3g^3 \over 16\p}\sqrt{{11 \over 6}} T^2 \nn \\
&&-{3\la T\over 2\p}
\sqrt{({3 \over 16}g^2+{\lambda \over 2}+{1 \over 4}f_t^2)(T^2-T_b^2)}\quad,
\eea
and the condition of vanishing curvature, $m_{\ph}^2(0,T'_b)=0$,
yields the corrected barrier temperature
\beq
T_b'^{-2}=\frac{1}{16\la v^2} \left(
{\displaystyle 3g^2+8\la+4f_t^2
-\frac{\sqrt{66}}{2\p}g^3}
+{\cal O}(g^4,\la^2,f_t^4)\right).
\eeq
Note, that there is no correction of order $\la^{3/2}$.

At $\ph>0$, the gap equations (\ref{gapeqml}) - (\ref{gapeqmx}) contain
terms proportional to $T/m_i$ which reflect the expected infrared
problems of finite-temperature perturbation theory.
The perturbative expansion is reliable if these terms
are smaller than leading-order terms.
Inspection of eqs. (\ref{gapeqml}) - (\ref{gapeqmx}) shows
that this is guaranteed if the following inequalities are satisfied:
\bea
\xi\ \frac{g^2}{6\p}\frac{T}{\mt+\emph}\leq 1\label{mboundsmt} \quad,\\
\xi\ \frac{\la T}{4\p}\left(\frac{3}{\emph}+\frac{1}{\emch}\right)
\leq 1 \quad.\label{mboundsmp}
\eea
The first inequality stems from the equation for $\mt^2$ and the second from
the equation for $\mph2$.
We have included a factor $\x$ which
ensures that leading terms of order $g^2$ and $\la$ are $\x$ times larger than
next-to-leading contributions.
Clearly, for fixed values of $g,\lambda,f_t,T$ and $\ph$,
the larger the value of $\x$,
the better the behaviour of the perturbative expansion.
If the conditions (\ref{mboundsmt}) and (\ref{mboundsmp}) are satisfied
one expects the uncertainty of the field dependent part of the
effective potential (\ref{thepotential1})
to be of order $1/\xi^2$.
We will now study the implications of the above conditions in two ways.
First we will use the iterative solution of the gap equations and
insert the squared plasma masses to order $g^2$ and $\lambda$
(cf. eq. (\ref{mass0})) into the conditions
(\ref{mboundsmt}) and (\ref{mboundsmp}).
We will then repeat the analysis in a more conservative way
by using the squared plasma masses in the $\xi$-conditions
to order $g^3$ and $\lambda^{3/2}$,
which are smaller than the ones to leading order
(cf. eq. (\ref{exactmass})).

Close to the origin, at $\ph\approx 0$, the above conditions imply
that one cannot even reach
the barrier temperature $T_b'$. For $\g=0$
eqs. (\ref{mass0}), (\ref{mboundsmt}) and
(\ref{mboundsmp}) yield the lower bounds on the temperature $T$,
\beq
T>T_V \quad, \qquad T>T_S \quad,
\eeq
where
\beq\label{tv}
\frac{{T_V}^2-T_b^2}{{T_V}^2}=\frac{\x_V^2g^4}{36\p^2
({3 \over 16}g^2+{\lambda \over 2}+{1 \over 4}f_t^2)}
\eeq
and
\beq\label{ts}
\frac{{T_S}^2-T_b^2}{{T_S}^2}=\frac{\x_S^2\la^2}{3\p^2
({3 \over 16}g^2+{\lambda \over 2}+{1 \over 4}f_t^2)}\,.
\eeq
Here, the subscript ``$V$'' (``$S$'') indicates that the infrared divergence
for the vector (scalar) field plasma mass sets the temperature.
Hence, our expression for the effective potential is only reliable for
temperatures
above $T^*$, which denotes the largest temperature among $T_b', T_V$ and
$T_S$. Non-zero values of $\g$ give a non-zero magnetic mass
$m_T$ at $\ph \approx 0$. The corresponding
infrared behaviour leads to a smaller temperature $T_V$.

In order to establish the existence of a first-order phase transition
for a given value of the scalar self-coupling $\lambda$ we have to
show that for some values $\ph>0$, where the $\xi$-conditions
(\ref{mboundsmt}) and (\ref{mboundsmp}) are satisfied,
\beq
V(\ph,T^*) \leq V(0,T^*) \,.
\eeq
Note, that $T^*=T^*(g,\lambda)$,
and thus the $\xi$-conditions are satisfied at $\ph=0$.
The larger the Higgs mass, i.e., the scalar self-coupling $\la$,
the more difficult it is to fulfill the condition for $\x_S$.
Let us now consider the potential in a region $\ph>0$, where the
$\xi$-conditions are satisfied.
At a certain maximal value of $\lambda$, and a corresponding critical
temperature $T^*_c$, the minimum of the potential
in this region, at $\ph_*>0$, is
degenerate with the minimum at $\ph=0$,
\beq
V(\ph_*,T^*_c)=V(0,T^*_c)\,.
\eeq
Note, that in general $T^*_c(g,\lambda)\leq T_c(g,\lambda)$, the true critical
temperature, and $\ph_*(g,\lambda)\geq \ph_c(g,\lambda)$, the local minimum
at $T_c$. Hence, $\ph_*$ is not necessarily an extremum of the potential
with respect to the full range in $\ph$. As discussed above,
the condition $T^*=T^*_c$ defines a maximal value of $\lambda$ for a
given value of $\xi$. We have plotted
the corresponding functions $m_H(\xi_S)$ and $m_H(\xi_V)$ in fig. \ref{xi}.
The condition obtained from the gauge boson gap equation is always
satisfied with $\x_V \geq 7.6$, even for $\gamma=0$.
The result is essentially independent of
the top mass in the range $m_t = 110 - 180\ GeV$.
Both, $\x_V$ and $\x_S$ are larger than 2
up to Higgs masses of approximately $200\ GeV$.

Let us now repeat this analysis using
the higher order plasma masses (\ref{exactmass})
in the $\xi$-conditions.
The allowed minimal temperature $T^*$
is higher than in the previous case. Since the Goldstone mass
$m_\chi^2 = 2{\partial V / \partial \ph^2}$ vanishes
at the second minimum $\ph(T^*)$, the condition (\ref{mboundsmp})
cannot be satisfied here and the perturbative approach breaks down.
Hence, one always has $\ph_*>\ph(T^*)$.
We can now determine the temperatures $T^*(g,\lambda)$.
In order to illustrate how far $T_c^*$ lies below the true
critical temperature $T_c$, we have plotted an
effective potential at $T_c^*$ in fig. \ref{effpot}.
The condition $T^*(g,\lambda)=T^*_c(g,\lambda)$ again defines
a line in the $\xi$-$\lambda$-plane, denoted by $\xi'_S$ in
fig. \ref{xi}. The difference between the two boundaries $\xi_S$
and $\xi'_S$ in fig. \ref{xi} is considerable.
For Higgs masses above
$80 \ GeV$ the value of $\xi'_S$ is smaller than $2$.

\bigskip

\centerline {\bf B. The effect of a magnetic mass}

\bigskip

Our estimate of the largest Higgs mass up to which perturbation theory
is reliable, which we have carried out in the previous subchapter, depends
on the parameter $\xi$. The appropriate value of $\xi$ can only be
determined by calculating higher order corrections. In sect. 4C we saw
that the magnetic mass has an important effect on the phase transition.
Hence, the effective potential has to be calculated at least up to
terms proportional to $g^6$, the order at which the magnetic
mass contributes. Since we have not calculated all graphs contributing
to this order, we have introduced a constant $\gamma$ in eqs. (\ref{mass0})
which parameterizes the uncertainty in the size of the magnetic mass.
The solution of our gap equations gives $\gamma=1$, other approaches
(cf. ref. \cite{kalas}) yield values of the same order of magnitude.

In order to estimate the effect of the magnetic mass on the phase transition
we calculate the surface tension, a physical quantity,
from the potential (\ref{thepotential1})
as function of $\lambda$ and $\g$:
\beq \label{s}
\sigma(\lambda,\g)=\int_0^{\ph_c}d\ph\sqrt{2V(\ph,T_c)},
\quad \ph_c=\ph(T_c).
\eeq
As discussed in the previous section, the temperature $T_c$ is always
larger than the temperature $T^*_c$, if the chosen value of $\lambda$
is allowed by the value chosen for $\xi$. Therefore one can reliably
calculate the potential close to the origin, $\ph\approx 0$, and
for large values of $\ph$. Of course, in the intermediate regime,
which is needed in order to get the surface tension from eq. (\ref{s}),
the potential has large uncertainties. Nevertheless, as the dominant
contribution comes from the vector loops which have no infrared
problems in the intermediate range of $\ph$, it is conceivable that
the surface tension, which is an integrated quantity, can be obtained
from eqs. (\ref{thepotential1}) and (\ref{s}) to good approximation.

We have plotted $\s$ as function of the Higgs mass for different
values of $\g$ in fig. \ref{surfacetension1}.
Note, that the disappearance of the surface tension means transition to
a second-order phase transition. The Higgs masses for which
$\s$ vanishes for different values of $\g$ are in agreement
with fig. \ref{gamma}.
For a given value of $m_H$ the effect of the magnetic mass on the phase
transition can be characterised by the ratio
\beq
\zeta(\gamma)=1-{|\s(m_H,\g)-\s(m_H,0)| \over \s(m_H,\g)+\s(m_H,0)} \quad.
\eeq
Clearly, if $\zeta(\gamma)\ll 1$
unknown higher order corrections are so large that a first-order
phase transition cannot be established unequivocally. According to
fig. \ref{surfacetension1} the quantity $\zeta$ is of ${\cal O}(1)$ for Higgs
masses
up to about $70 \ GeV$.
This value of $m_H$ is rather close to the present
lower experimental bound.

\bigskip

\centerline {\bf C. Scalar loop and superdaisy contributions}

\bigskip

In this subchapter we analyse the role of some higher order terms in
the perturbative expansion. As in the previous
subchapter we again study the surface tension (\ref{s}).

It is instructive to first set $g=0$ and to look at the perturbative
expansion in $\lambda$. The phase transition of the pure scalar theory
is well known to be second-order.
Our results agree with this fact, since the effective potentials
in the
leading order, ${\cal O}(\lambda)$, and the next-to-leading
order, ${\cal O}(\lambda^{3/2})$,
give second-order phase transitions.
The appearance of a first-order phase transition
would only be a sign that our perturbation expansion
is not reliable, since it would result from a cancellation between
terms of different orders. In other words,
contributions from different orders would be of the
same order of magnitude.

Let us now study the case $g\neq 0$. The scalar contributions together with
the term ${\cal O}(g^2)$ give a second-order phase transition,
while inclusion of
terms of order $g^3$ yields a first-order transition.
The above mentioned problem is not relevant here,
since the cancellation takes place between terms proportional
to different expansion parameters, namely $g^3$ and $\lambda$.
In this perturbative approach, where $g^3\approx \lambda$,
the size of the
next-order terms (in our case $\lambda^{3/2}$
and $g^4$) must be small, and in particular their effect
on the surface tension has to be a reasonable correction.

We have determined the contribution ${\cal O}(\lambda^{3/2})$ to the
effective potential.
In fig. \ref{surfacetension2} we have plotted
the surface tension at the critical temperature
as function of the Higgs mass for the potential
containing terms ${\cal O}(\lambda^{3/2},g^3)$ (solid line) and
for the potential with terms ${\cal O}(\lambda,g^3)$ (long dashed line).
For small Higgs masses the difference is small, but
at approximately $100 \ GeV$ the difference becomes ${\cal O}(50\%)$
indicating a breakdown of the perturbative approach.

Superdaisy diagrams can be summed by finding exact solutions to the
gap equations. Iterating the gap equations
we have determined the contribution of the superdaisy
diagrams to order $g^4$ and $\lambda^2$.
The rather lengthy expressions for the potential are given in appendix B.
In fig. \ref{surfacetension2} we have plotted
the surface tension at the obtained critical temperature
for this effective potential (short dashed line). The superdaisy
terms give a correction ${\cal O}(20\%)$ correction, and thus do not change
the qualitative picture of a first-order phase transition.

\section{Decay of metastable states}\label{droplets}

In condensed matter physics the decay of metastable states is described by
Langer's theory \cite{wb16}. The starting point is a coarse-grained
free energy which depends on the order parameter, temperature and the coarse
graining scale. As function of the order parameter the free energy has a
local metastable minimum which is separated from the global minimum by
a barrier whose height determines the lifetime of the metastable state.
{}From a stationary solution of the Fokker-Planck equation
for the probability distribution of large fluctuations of the order parameter
(``subcritical droplets'') one then obtains a formula for the decay
rate \cite{wb16} which depends on the free energy of the metastable state,
the free energy of a saddle point field configuration which interpolates
between local and global minimum and a "dynamical factor" which cannot
be obtained from equilibrium thermodynamics.

Langer's formalism can be directly applied to the decay of metastable states
in quantum field theories. The case of scalar electrodynamics has been studied
in a recent paper \cite{u1}. The decay rate is given by
\beq\label{ourrate}
\G={\k\0 2\p}\,\frac{\mbox{Im} \zb[\bar{\PH}]}{\zb[\PH=0]} \quad,
\eeq
where
\bea\label{zbeta}
\zb[\PH]&=&e^{\textstyle -\be F[\PH,T]}\\
        &=&\int_\be[D\hat{\PH}][DA_\m][D\psi]
        e^{\textstyle -\left(S_\be[\PH+\hat{\PH},A_\m,\psi]
        -\int_\be dx \frac{\de F[\PH,T]}{\de \PH(x)}\hat{\PH}(x)\right)}.
\nn
\eea
The integration measure for the vector field includes the gauge fixing and
ghost terms, and $\psi$ stands for all fermion fields.
In general, the free energy also depends on $A_\mu$ and $\psi$. However,
we will only consider stationary points with $A_{\mu} = \psi = 0$.
For simplicity we have therefore omitted the dependence of $F[\Phi,T]$
on $A_{\mu}$ and $\psi$ as well as the corresponding functional derivative
terms in the exponent of eq. (\ref{zbeta}).
$\bar{\Phi}$ is a field configuration which interpolates between the
symmetric and the broken phase. Since $\Phi=0$ and $\bar{\Phi}$ are approximate
stationary points of
the free energy $F[\Phi,T]$, we neglect
the second term of the integrand in eq. (\ref{zbeta}).

The functional integral over vector and fermion fields
yields an effective action which
depends on the scalar field $\Phi+\hat{\Phi}$,
\bea
&&\int_\be[DA_\m][D\psi]
        e^{\textstyle -S_\be[\PH+\hat{\PH},A_\m,\psi]}\nn\\
&&\quad=\exp\big(-\int_{\be}dx (\partial_{\m}(\PH+\hat{\PH})^{\dagger}
\partial^{\m}(\PH+\hat{\PH})-V_0(z)\nn\\
&&\hphantom{\quad\exp\big(-\int_\be}- \tilde{V}(z,T)
+ \tilde{Z}(z,T)\partial_\m(\PH+\hat{\PH})^{\dagger}\partial^\m
(\PH+\hat{\PH})+\dots)\big),
\eea
where $z=\sqrt{2(\PH+\hat{\PH})^{\dagger} (\Phi+\hat{\Phi})}$.
Note, that this expression is invariant under the global symmetry
$O(4)$ rather than $SU(2)\times U(1)$ which is the
symmetry of the Lagrangian given in eq. (\ref{theory}). Indeed,
integrating out the top-quark yields additional wave function correction
terms $\tilde{Z}_i(\Phi+\hat{\Phi})$ which are invariant only under the
smaller symmetry group. For simplicity
we will neglect all wave function correction terms in the following.
A more detailed discussion will be given elsewhere \cite{wb21}.

At one-loop order the potential $\tilde{V}(z,T)$ is well known.
Including vector boson and top quark loops one obtains for the
full potential in the high temperature expansion
(cf. eq.(\ref{thepotential})):
\bea \label{prepotential}
\bar{V}(\ph,T)&=&V_0(\ph) + \tilde{V}(\ph,T) \nn\\
              &=&{1 \over 2}(\frac{3g^2}{16}+{1 \over 4}f_t^2)
(T^2-T_b^2)\ph^2+{\la\over 4}\ph^4\nn\\
        & &- \frac{3 g^3}{32\pi}\ph^3 T + {\cal O}(g^4,f_t^4).
\eea
The potential has two local minima, at $\ph=0$ and at $\ph(T)>0$. At
the critical temperature $T_c$ both minima are degenerate and one has
\bea \label{tc}
\ph_c \equiv \ph(T_c) = \frac{g^3}{8\pi\lambda}T_c .
\eea

The integral over the scalar field fluctuations $\hat{\PH}$ can now be carried
out in the saddle point approximation.
In the thin wall approximation \cite{wb18} the stationary point of the
approximate free energy
\beq\label{vsuba}
\bar{F}[\PH,T]=\int d^3x\,\big(|\vec{\nabla}\PH|^2+\bar{V}(z,T)\big)\quad,
\eeq
which appears in the integrand of eq. (\ref{zbeta}),
has a saddle point $\bar{\Phi}$ which interpolates between $\Phi=0$
and $\Phi(T)>0$. For temperatures just below the critical temperature $T_c$
the height of the barrier between the two minima is large compared to the
potential difference between the minima. In this case the saddle point
can be computed in the thin wall approximation and one obtains
\beq\label{saddle}
\bar{\PH}(r)=\frac{1}{\sqrt{2}}\bar{\ph}(r)
=\frac{1}{2\sqrt{2}}\ph_c\Big[1-\tanh\Big(\frac{r-R(T)}{d}\Big)\Big],
\eeq
with
\bea\label{dvont}
d&=&\frac{2\sqrt{2}}{\sqrt{\la}\ph_c},\nn\\
\s&=&\int_0^{\ph_c}d\ph\ \sqrt{2\bar{V}(\ph,T_c)},\nn\\
R(T)&=&\frac{2\s}{\varepsilon(T)},\quad
\varepsilon (T)= \bar{V}(0,T)-\bar{V}(\ph(T),T).
\eea
The free energy $\bar{F}_{TW}[\bar{\PH},T]$ in this approximation
is then the sum of a volume term and a
surface term:
\beq\label{surfandvol}
\bar{F}_{TW}[\bar{\PH},T]=4\p R^2(T)\s-{4\p\0 3}R^3(T)\varepsilon(T).
\eeq
It is sufficient to evaluate the surface tension $\s$ at the critical
temperature $T_c$.

For the partition function of the saddle point one now obtains
\beq
\mbox{Im}\ \zb[\bar{\Phi}]={1\0 \sqrt{|\bar{\la}_{-}|}}
{\cal V}{\prod_i}^{>}\,(\bar{\la}_i)^{-1/2}\,
e^{\textstyle -\be \bar{F}[\bar{\Phi},T]},
\eeq
where ${\prod}^>$ denotes the product of all
positive eigenvalues $\bar{\la}_i$ of
fluctuations around $\bar{\Phi}$, $\bar{\la}_{-}$ is the single negative
eigenvalue, and ${\cal V}$ is the volume of zero modes associated with the
symmetries of the system under consideration.

The scalar fluctuations $\hat{\PH}$ consist of the radial modes $\hat\ph$
and the Goldstone modes $\hat{\chi}_i$:
\bea
\bar{F}[\bar{\PH}+\hat{\PH},T]&=&\bar{F}[\bar{\PH},T]\nn\\
&&+\2\int_\be dx\ \hat{\ph}(x)
(-\triangle+U_\ph(r))\hat{\ph}(x)\nn\\
&&+\2\int_\be dx\ \hat{\chi}_i(x)
(-\triangle+U_\chi(r))\hat{\chi}_i(y).
\eea
The corresponding potentials for the scalar fields $\hat{\ph}$ and $\hat{\chi}$
are
\bea
U_\ph(r)&=&
\frac{\6^2}{\6\ph^2}\,\bar V(\ph,T)\Big|_{\ph=\bar{\ph}(r)},\\
U_\chi(r)&=&
\frac{1}{\ph}\frac{\6}{\6\ph}\,\bar V(\ph,T)\Big|_{\ph=\bar{\ph}(r)}.
\eea
The spectrum of eigenvalues contains six zero modes,
three for translational invariance and
three for the global $SU(2)\times U(1)$ symmetry of $\bar{F}[\PH,T]$
which is spontaneously broken to the electromagnetic $U(1)$ subgroup.
The corresponding volume factor for the translational modes is well
known. The volume factor for the global symmetry can be calculated in
the same way as for sphaleron tunneling processes, the details are given
in appendix C. One obtains:
\bea\label{volume}
{\cal V}_\ph&=&\left({\be\0 2\p}\bar F_A[\bar{\PH},T]\right)^{3/2}V,\nn\\
{\cal V}_\chi&=&\frac{\pi^2}{2}
\left({\be\0 2\p}\int d^3x\,\bar{\ph}^2\right)^{3/2},
\eea
where $V$ is the total volume of the physical three dimensional space.

The discrete $\hat{\ph}$ spectrum is well known \cite{wb16,wb19}, since the
bound states are localized at $r\approx R$. There is one negative eigenvalue,
\beq\label{lamin}
\la_{-}\approx -{2\0 R^2},
\eeq
which guarantees that $Z_{\be}[\bar{\PH}]$ is purely imaginary. Furthermore,
there are ``Goldstone modes''  which correspond to
deformations of the droplet surface \cite{wb16,wb19,wb20}.
The corresponding contribution to the determinant of eigenvalues
is $(\m R)^{-5/3}$, where $\m=\emph(0,T)$ \cite{buhel}.
Combining eqs. (\ref{ourrate}), (\ref{surfandvol}), (\ref{volume}) and
(\ref{lamin}) we finally arrive at the transition rate
\beq\label{finalrate}
\frac{\G}{V}=\frac{\sqrt{2}}{2^{9}\cdot 3^3\cdot \pi^2}
\frac{g^9}{\la^3}\,\k\,
\left(\beta\sigma\right)^{3/2}
\left(\beta\m\right)^{-3/2}
(R\mu)^{41/6}\, e^{\textstyle -{4\p\0 3}\be\s R^2}.
\eeq
Here the contributions of zero modes and Goldstone modes to the
determinant of scalar fluctuations around the saddle point have been taken into
account. The "dynamical factor" $\kappa$ has recently been evaluated
\cite{Csernai},
\beq
\kappa = \frac{16\eta\sigma}{3(\Delta\omega)^2 R^3}\,,
\eeq
where $\Delta\omega$ is the difference of the enthalpy
$\omega=-T\partial V/ \partial T$ between symmetric and broken phase,
and the viscosity $\eta=65.4\ T^3$ in the standard model \cite{Csernai}.
The pre-factor of the exponential in eq. (\ref{finalrate})
and the naive estimate $T^4$ turn out to give
in our numerical calculations
essentially the same results.
However, a priori this is not clear, and
the evaluation of the pre-factor is needed to verify
the validity of the semiclassical approximation.

Eq. (\ref{finalrate}) gives the decay rate of the metastable symmetric phase
in the framework of Langer's theory of metastability. The free energy
$\bar{F}[\PH,T]$,
obtained by integrating out vector boson and fermion fields,
plays the role of the
coarse-grained free energy in condensed matter physics. An
important aspect of this approach is that scalar fluctuations are only computed
around the stationary points $\PH=0$ and $\PH=\bar{\PH}$ of $\bar F_A[\PH,T]$
and
\underline{not}, as usually done, around unstable homogeneous scalar
background fields. Hence, the perturbative approach
is consistent and does not break
down due to infrared divergencies or negative scalar mass terms.
The decay rate (\ref{finalrate}) is similar to the result obtained for
scalar electrodynamics \cite{u1}. The difference in the pre-factor is due
to the different global symmetries which are spontaneously broken.
\renewcommand{\arraystretch}{1.5}
\begin{table}[bt]
\begin{center}
\begin{tabular}{l|rrrr}
\hline
&
$\bar{V}$&$V_{2/3}$&$V_\lambda$&$V_\gamma$\\
\hline
$T_c$ $[\mbox{GeV}]$&
$102$&$102$&$103$&$103$\\
$T_c-T_e$ $[\mbox{GeV}]$&
$0.05$&$0.02$&$0.02$&$0.01$\\
$T_c-T_b$ $[\mbox{GeV}]$&
$0.63$&$0.28$&$0.36$&$0.05$\\
$\ph_c/T_c$&
$0.38$&$0.25$&$0.25$&$0.23$\\
$\sigma$ $[\mbox{GeV}^3]$&
$1396$&$409$&$624$&$187$\\
$F/T$&
$240$&$223$&$232$&$221$\\
$F_{TW}/T$&
$143$&$143$&$143$&$143$\\
$R$ $[\mbox{GeV}^{-1}]$&
$1.6$&$2.9$&$2.4$&$4.3$\\
$R/\xi$&
$8$&$10$&$9$&$6$\\
$d/R$&
$0.30$&$0.25$&$0.28$&$0.29$\\
\hline
\end{tabular}
\caption{\label{Tab1}
Observables of the first-order phase transition
for four different effective potentials,
a Higgs mass of $70$ GeV and a top-quark
mass of $140$ GeV.
}
\end{center}
\end{table}

Let us finally consider the cosmological phase transition. A rough
estimate of the temperature $T_e$
at which the phase transition ends, is obtained by requiring
\beq\label{te}
\G(t_e)t_e^4\sim 1,
\eeq
where $t\approx 0.03\cdot m_{pl}/T^2$ .
As an example we choose for Higgs boson and top quark masses the
values $m_H=70\ GeV$ and $m_t=140\ GeV$. The W-boson mass is $m_W=80.6\ GeV$.
These masses correspond to the coupling constants $\la=0.04,\ f_t=0.80$
and $g=0.66$. From our discussion in sect. 5 we know that for these
parameters higher order corrections are under control so that we
can still say that the phase transition is first-order. From the
free energy $\bar{F}$ (cf. eq.(\ref{vsuba})) we can compute the critical
temperature $T_c$, the barrier temperature $T_b$,
the surface tension $\sigma$, the correlation
length of the symmetric phase $\xi= 1/ \mu$, and the value of
the Higgs field inside the droplet, i.e. ${\ph_c / T_c}$.
{}From eqs. (\ref{finalrate})
and (\ref{te}) we then obtain the temperature $T_e$ and the size of
$R$ of the critical droplet. All these quantities are listed in
table 1. Note, that the size of the pre-factor in the total
rate relative to the exponential is of order $1\%$. The thin wall
approximation for the Higgs mass used in the table is marginally applicable.
We have plotted $\bar V$ on
fig. \ref{pend} for three different Higgs masses at the corresponding
temperature $T_e$. As it can be seen, the larger the Higgs mass the better the
thin wall approximation.

How reliable are these results?  Since the ratio
${\ph_c / T_c}$ is rather small, the longitudinal plasma
mass $m_L$ (cf. eq. (\ref{gapeqml})) is essentially independent of $\ph$.
Hence, the longitudinal degree of freedom of the vector field
decouples and does not contribute to the tunneling process. This leads
to a reduction of the cubic term in the effective potential by ${1\over 3}$
\cite{wb6}. This effect is not included in the potential $\bar{V}$. Let
us denote the potential with the reduced cubic term by $V_{{2\over 3}}$.
As shown in table 1, the effect on the surface tension is considerable.
However, the phase transition clearly remains first-order with some
change in the relevant temperatures and droplet properties. In order
to illustrate the possible effect of higher order corrections we
have also computed the observables for the potential including the
scalar loops ($V_{\la}$) and for the potential with scalar loops and a
non-vanishing magnetic mass ($\gamma=1$) (cf. eq. (\ref{thepotential1})).
Note in particular the effect of the magnetic mass on the surface
tension. A more detailed quantitative discussion will be presented
in a separate paper \cite{wb21}.

To conclude, we have obtained a consistent description of a cosmological
first-order electroweak phase transition for values of the Higgs boson
and top-quark masses which are compatible with present experimental limits.
The phase transition is only weakly first order. The quantitative description
of the transition can be used as input for models of baryogenesis.

\section{Summary}\label{summary}
In the previous sections we have studied the transition from the symmetric to
the broken phase in the $SU(2)$ gauge theory at finite temperature.
We have seen
that, due to infrared divergencies, ordinary perturbation theory to any finite
number of loops does not yield a useful approximation
to the effective
potential. However, an improved perturbation theory, which takes plasma masses
into account, describes consistently the symmetric phase ($\ph=0$)
and also the broken phase ($\ph>0$) in the neighbourhood of the second
non-trivial, local minimum of the effective potential. Using this improved
perturbation theory we have evaluated the effective potential including all
terms cubic in the couplings and shown that
all contributions linear in $\ph$ cancel. To this order in the couplings
the ring summation and the improved perturbative approach are equivalent.

The plasma masses have been determined from a set of one-loop gap equations.
A non-vanishing  transverse gauge boson  plasma mass was obtained,
originating from the non-abelian gauge interactions.
Based on the gap equations
we also found a range in the couplings $g$, $\la$ and $f_t$,
the temperature $T$
and the scalar field $\ph$, where the perturbative approach is reliable.
The dependence on the top-quark Yukawa coupling
$f_t$ turns out to be irrelevant. Knowing this
range in $T$ and $\ph$ as function of $g$ and $\la$ where the effective
potential is reliable has allowed us to determine the range in $\la$
where the symmetric phase is metastable.
As a criterion we required that at the origin, $\ph=0$, the effective
potential
has only a local and not a global minimum for the allowed values of $T$.

We find that the electroweak phase transition is weakly first-order
and that our perturbative approach in fact breaks down for Higgs masses
close to the present experimental lower bound. Higher order corrections
from scalar loops become important for Higgs masses around $80\ GeV$, for
vanishing magnetic mass of the $W$-boson (cf. sects. 4A and 4C). For
non-vanishing magnetic mass ($\gamma=1$) the perturbative approach breaks down
around $m_H=70\ GeV$.

Following the theory of Langer we have finally computed the nucleation rate for
critical droplets, and we have discussed some aspects of the
cosmological phase transition. Up to Higgs masses of order $80\ GeV$
the picture of a first-order transition, which proceeds via nucleation
and growth of critical droplets, appears self-consistent,
if the magnetic plasma mass vanishes. However,
the thin wall approximation is only marginally applicable. Furthermore,
the Higgs vacuum expectation value inside the critical droplet is
much smaller than the value required by models of electroweak baryogenesis.

Our results could be improved in several respects. Clearly,
a complete computation of the
effective potential to order $g^4, \lambda^2$ and $f_t^4$
would be very valuable in order to further test the convergence of the
perturbation theory. A method to identify the relevant
contributions is given in appendix A.
Furthermore, the validity of the expansion in powers of derivatives used in
sect. \ref{droplets} has to be examined in greater detail.
However, the most crucial ingredient concerning the order of the transition
and the difference between abelian and non-abelian gauge theories appears
to be the magnetic gauge boson mass, whose origin and size require
further investigations.

We would like to thank
M. L\"{u}scher, I. Montvay, M. Reuter, N. Tetradis and C. Wetterich for
helpful discussions and comments.

\vfill\eject

\appendix{\bf Appendix A \, Ring diagram summation}
\renewcommand{\theequation}{\mbox{A}.\arabic{equation}}
\setcounter{equation}{0}

\bigskip

As it was promised we will now prove that only ring diagrams give
non-zero contributions cubic in the couplings
($g, \sqrt{\lambda}$ and $f_t$).

Instead of $g, \sqrt{\lambda}$ and $f_t$ we will use in the
following a generic coupling $h$. The naive order of a
Feynman diagram is given by
\beq
{\cal O}_{naive}=V^{(1)}_{naive}+2V^{(2)}_{naive},
\eeq
where
$V^{(1)}_{naive}$ is the number of the vertices of order $h$
(the triple vector vertex, the ghost-ghost-vector vertex,
the scalar-scalar-vector vertex and the fermion-fermion-boson vertices)
and $V^{(2)}_{naive}$ is the number of the vertices of order
$h^2$ (the scalar triple self-coupling, the vector-vector-scalar
vertex and the quartic boson vertices).
We study a graph with non-zero number of vertices contributing
to the effective potential.
It is one-particle irreducible with no external lines. Momentum
conservation gives a relationship between the total number of
vertices ($V=V^{(1)}_{naive}+V^{(2)}_{naive}$), the number of independent
loops ($L$) and internal lines ($I$),
\beq\label{momentum}
V+L-I=1.
\eeq

The momenta of the internal lines $q_1,q_2,...,q_I$ are linear
combinations of the loop variables $p_1,p_2,...,p_L$.
A lower order contribution in $h$ (${\cal O}<{\cal O}_{naive})$
could arise if some
of the Matsubara frequencies, e.g. $q_{10},q_{20},...q_{s0}$,
vanish. Denote the loop momenta with vanishing Matsubara
frequencies as $p_1,p_2,..,p_r$.
We call lines or loops with vanishing Matsubara frequencies
{\underline {soft}}, those
with non-vanishing Matsubara frequencies {\underline {hard}} (cf. ref.
\cite{Pisarski}).
Fermions have always non-zero Matsubara
frequencies, thus fermion
lines and loops are always hard.

The substitution in the soft loop variables
\beq
\vec{p}_1=m_1^{(L)}\vec{y}_1 \ ,\ \ \ \vec{p}_2=m_2^{(L)}\vec{y}_2
\ ,\ \ \ ... \ \ \ ,\  \vec{p}_r=m_r^{(L)}\vec{y}_r
\eeq
results in a similar substitution for the propagators
\beq
\vec{q}_1=m_1^{(I)}\vec{x}_1 \ ,\ \ \ \vec{q}_2=m_2^{(I)}\vec{x}_2
\ ,\ \ \ ... \ \ \ ,\  \vec{q}_s=m_s^{(I)}\vec{x}_s.
\eeq
Here the masses are proportional to the coupling $h$
\bea
m_i^{(L\ or\ I)}=ha_i^{(L\ or\ I)},\ \
a_i= \cases  {&$\ph/2$\,  for\  vector boson; \cr
&$\sqrt{\ph^2-v^2}$\,  for\  Goldstone bosons; \cr
&$\sqrt{3\ph^2-v^2}$\,  for\  the Higgs boson\,.\cr}
\eea

The above transformation gives for each soft loop integration,
\beq
\int d^3p_i=m_i^3\int d^3 y_i \propto h^3\int d^3 y_i\,,
\eeq
an additional factor of $h^3$, and for each soft boson propagator,
\beq
D({\vec q}_i)={1 \over m_i^2}D({\vec x}_i) \propto h^{-2}D({\vec x}_i)\,,
\eeq
an additional factor of $h^{-2}$. The same factorization can be done
for vertices with momentum dependence
if all of the incoming lines are soft,
\beq
{\cal W}({\vec p}_i,{\vec p}_j,{\vec p}_k) \propto h
{\cal W}({\vec x}_i,{\vec x}_j,{\vec x}_k)\,,
\eeq
where ${\cal W}$ is a linear combination of the soft momenta. Note, that
for the ghost-ghost-vector vertex this additional factor $h$
appears not only in the case
where all of the lines are soft, but also if one of the
incoming ghost lines are soft.

Hence, inspection of the Feynman rules suggest that the
diagram is at least of the order
\beq \label{order}
{\cal O}=2V^{(2)}+V^{(1)}+3r-2s,
\eeq
where $V^{(1)}$ is the number of the vertices with naive order $h$ and at
least two incoming hard lines.

Remove now the $r$ soft lines from the graph and consider the remaining
subdiagram containing only hard lines. This subdiagram is
not necessarily connected but all of the
$j=1,...,n$ connected parts are closed
graphs with no external lines. (The original graph has no
external lines and due to the momentum conservation at the
vertices there is no way to connect a hard line only to soft lines.)
The relationship (\ref{momentum}) is valid for each of these connected
graphs with $V_j$ vertices, $L_j$ loops and $I_j$ lines, thus
\beq\label{disconn}
V_j+L_j-I_j=1,\ \ j=1,..,n \ \ .
\eeq
Note, that in this case some of the vertices are connected only to two lines
but this fact has no influence on our consideration. Clearly,
the number of the vertices in the j-th connected graph $V_j$
is the sum of the number of the vertices of order $h$ and the number
of the vertices of $h^2$, thus
\beq\label{v0}
V_j=V_j^{(1)}+V_j^{(2)}.
\eeq
The sum of the hard lines in the individual connected hard subdiagrams
is the total number of the hard lines $(I-s)$,
and the sum of the individual hard loops is the
total number of the hard loops $(L-r)$,
\beq
\sum_j L_j=L-r, \ \ \ \sum_j I_j=I-s.
\eeq
Since the vertices of order $h$ have at least two hard lines the
connected hard subdiagrams contain all of them
\beq\label{v1}
\sum_j V_j^{(1)}=V^{(1)}.
\eeq
Summing (\ref{disconn}) over $j$ one obtains
\beq
\sum_j(V_j-1)+(L-r)-(I-s)=0.
\eeq
Solving for $s$ and inserting it into (\ref{order}) yields
\bea
&&{\cal O}=2V^{(2)}+2V^{(1)}-V^{(1)}+r+2r-2s \nn \\
&&=2V-V^{(1)}+r+2r+2\left(L-r-I-\sum_j(V_j-1)\right).
\eea
{}From this equation, the relation $V+L-I=1$, (\ref{v0}) and (\ref{v1})
one obtains the final answer
\beq
{\cal O}=(2+r)+\sum_j(V_j^{(1)}+2V_j^{(2)}-2).
\eeq

The case $r=0$ corresponds to no soft lines, thus the order of the
diagram is just the naive order.
For $r=1$ the order of the
diagram is at least $3$, where we have one soft loop. If
$V_j^{(1)}+2V_j^{(2)}-2=0$ for $j=1,...,n$, then the order of the
graph is still $3$. To ensure this for a given $j$ there are two possibilities.
Either there is only one quartic coupling of order $h^2$
in this hard subgraph or
there are two triple vertices of order $h$ (since we only study
one-particle irreducible contributions to the effective potential
it is not possible to have only
one triple vertex in a hard subgraph.) These are just the ring diagrams.
Note, that the  self-energy insertion with two scalar triple
vertices or vector-vector-scalar vertices do not belong to this
class.

If one wants to collect the graphs of order $h^4$ one has to
take into account the previous ($r=1$)
graphs one order further in the high temperature expansion,
a finite
number of graphs with no soft loops ($r=0$), no new terms with
one soft loop ($r=1$) and the above mentioned hard graphs
($V_j^{(1)}+2V_j^{(2)}-2=0$)
for the case with two soft loops ($r=2$). As before,
self-energy insertions to order $h^2$ are satisfactory.
\vfill\eject

\appendix{\bf Appendix B \, Partial summation of superdaisy graphs}
\renewcommand{\theequation}{\mbox{B}.\arabic{equation}}
\setcounter{equation}{0}
\bigskip

As we have shown performing an iterative solution of the gap equations
one can obtain higher order contributions to the effective potential.
In this appendix we calculate the $g^4$, $\lambda^2$ and
$f_t^4$ contributions.
Clearly, not all of this corrections are involved in the solution of
the gap equations, because a summation where all
the self-energy insertions are calculated
at zero external momenta cannot be justified.

The leading order masses (\ref{mass0}) were obtained by one iteration
of the gap equations, keeping the terms of order $g$,
$\sqrt \lambda$ or $f_t$. Inserting these masses in the gap equations
and performing a Taylor expansion up to order $g^2$, $\lambda$
and $f_t^2$ in the square-roots of the obtained  quantities, one gets
the next order masses.
\bea
m_L^{(2)}&=&
ga+\left (-{\frac {3\,d}{32\,a}}-{\frac {c}{32\,a}}-{\frac {b}{2\,a}
}-{\frac {\ph^{2}}{32\,a\left (c+a\right )}}\right )g^{2}T\pi ^{-1},
\nn \\
m_T^{(2)}&=&
gb+\left ({\frac {d^{2}}{96\,b\left (d+c\right )}}-{\frac {dc}{48\,b
\left (d+c\right )}} \right. \nn \\
&&
\left.
+{\frac {c^{2}}{96\,b\left (d+c\right )}}-{\frac {
\ph^{2}}{48\,b\left (c+b\right )}}\right )g^{2}T\pi ^{-1}, \nn \\
m_\ph^{(2)}&=&
gc+\left (
-{\frac {3\,\ph^{2}}{64\,cb}}-{\frac {3\,b}{16\,c}}-{\frac {3\,a}{32
\,c}}-{\frac {3\,\ph^{2}}{128\,ca}}\right )g^{2}T\pi ^{-1} \nn \\
&&
+\left (-{
\frac {3\,d}{8\,c}}-3/8\right )T\lambda\pi ^{-1} \nn \\
&&+\left (-{\frac {3\,
\ph^{2}}{8\,dc}}-{\frac {9\,\ph^{2}}{8\,c^{2}}}\right )\lambda^{2}T
\pi ^{-1}g^{-2}, \nn \\
m_\chi^{(2)}&=&
gd+\left (-{\frac {3\,b}{16\,d}}-{\frac {3\,a}{32\,d}
}\right )g^{2}T\pi ^{-1}+\left (-{\frac {c}{8\,d}}-5/8\right )T
\lambda\pi ^{-1} \nn \\
&&-{\frac {T\lambda^{2}\ph^{2}}{2\,dg^{2}\pi \,\left (d
+c\right )}}.
\eea

In order to reduce the size of our formulas $a,b,c$ and $d$ were introduced.
They are proportional to the leading order masses. Thus,
\bea
a&=&{m_L^{(1)} \over g}=\sqrt{{11 \over 6}T^2+{\ph^2 \over 4}}
\nn \\
b&=&{m_T^{(1)} \over g}={\ph \over 2}
\nn \\
c&=&{m_\ph^{(1)} \over g}
=\sqrt{({3 \over 16}+{\lambda \over 2g^2}+{f_t^2 \over 4g^2})(T^2-T_b^2)
+3{\lambda\ph^2 \over g^2}}
\nn \\
d&=&{m_\chi^{(1)} \over g}
=\sqrt{({3 \over 16}+{\lambda \over 2g^2}+{f_t^2 \over 4g^2})(T^2-T_b^2)
+{\lambda\ph^2 \over g^2}}.
\eea

By use of this masses the next iteration with a Taylor expansion of
order $g^4$, $\lambda^2$ and $f_t^4$ can be carried out.
After this straightforward but fairly tedious calculation
the next order Goldstone mass term is of the form:

\bea
 {m_\chi^{(3)}}^2&=&
\left (-{\frac {\ph^{4}}{2\,\left (d+c\right )^{3}d}}-{\frac {9\,\ph
^{4}}{8\,\left (d+c\right )^{2}c^{2}}}-{\frac {3\,\ph^{4}}{8\,\left (
d+c\right )^{2}cd}}\right )\lambda^{4}T^{2}\pi ^{-2}g^{-4} \nn \\
&&+d^{2}g^{2}
+\left (-{\frac {\ph^{2}c}{8\,\left (d+c\right )^{2}d}}-{\frac {\ph^
{2}}{\left (d+c\right )^{2}}}+{\frac {5\,\ph^{2}}{8\,d\left (d+c
\right )}} \right. \nn \\
&&\left.
+{\frac {9\,\ph^{2}}{32\,c^{2}}}-{\frac {3\,\ph^{2}d}{8\,
\left (d+c\right )^{2}c}}+{\frac {3\,\ph^{2}}{32\,cd}}\right )\lambda
^{3}T^{2}\pi ^{-2}g^{-2} \nn \\
&&+\left ({\frac {15\,b}{64\,d}}+{\frac {3\,a}{
128\,c}}+{\frac {3\,b}{64\,c}}+{\frac {3\,\ph^{2}}{256\,cb}}+{\frac {
3\,\ph^{2}}{512\,ca}}+{\frac {15\,a}{128\,d}}\right )\lambda\,T^{2}g^
{2}\pi ^{-2} \nn \\
&&-{\frac {T\lambda^{2}\ph^{2}}{g\pi \,\left (d+c\right )}}
+{\frac {\left (-{\frac {c}{4}}-{\frac {5\,d}{4}}\right )\lambda\,Tg}
{\pi }}+{\frac {\left (-{\frac {3\,a}{16}}-{\frac {3\,b}{8}}\right )T
g^{3}}{\pi }} \nn \\
&&+\left (-{\frac {d^{2}}{256\,b\left (d+c\right )}}+{
\frac {3\,b}{32\,a}}+{\frac {9\,d}{512\,a}}+{\frac {cd}{128\,b\left (d
+c\right )}}-{\frac {c^{2}}{256\,b\left (d+c\right )}}
\right. \nn \\
&&\left.
+{\frac {3\,c}{
512\,a}}+{\frac {3\,\ph^{2}}{512\,a\left (c+a\right )}}+{\frac {\ph^
{2}}{128\,b\left (c+b\right )}}\right )T^{2}g^{4}\pi ^{-2} \nn \\
&&+\left (-{
\frac {3\,\ph^{2}a}{32\,\left (d+c\right )^{2}c}}-{\frac {3\,\ph^{2}
b}{16\,\left (d+c\right )^{2}d}}-{\frac {3\,\ph^{4}}{128\,\left (d+c
\right )^{2}ca}}
\right.
\nn \\
&&\left.
-{\frac {3\,\ph^{2}a}{32\,\left (d+c\right )^{2}d}}
-{\frac {3\,\ph^{4}}{64\,\left (d+c\right )^{2}cb}}-{\frac {3\,\ph^{2}
b}{16\,\left (d+c\right )^{2}c}}+{\frac {7}{8}}+{\frac {5\,c}{32\,d}}
\right.
\nn \\
&&\left.
+
{\frac {3\,d}{32\,c}}\right )T^{2}\lambda^{2}\pi ^{-2}.
\eea

Similarly, it is easy to get the corresponding Higgs mass-squared of order
$g^4$, $\lambda^2$ and $f_t^4$.

\bea
{m_\ph^{(3)}}^2&=&
g^{2}c^{2}+\left (-{\frac {27\,\ph^{4}}{512\,c^{3}a}}+3/4-{\frac {9
\,\ph^{2}b}{64\,d^{3}}}-{\frac {9\,\ph^{2}a}{128\,d^{3}}}
\right. \nn \\
&&
\left.
-{\frac {27\,\ph^{2}a}{128\,c^{3}}}
-{\frac {27\,\ph^{2}b}{64\,c^{3}}}+{\frac {9
\,d}{32\,c}}+{\frac {3\,c}{32\,d}}-{\frac {27\,\ph^{4}}{256\,c^{3}b}}
\right )T^{2}\lambda^{2}\pi ^{-2} \nn \\
&&
+\left ({\frac {9\,\ph^{2}}{256\,cb}
}+{\frac {9\,a}{128\,c}}+{\frac {9\,b}{64\,d}}+{\frac {9\,b}{64\,c}}+{
\frac {9\,\ph^{2}}{512\,ca}}+{\frac {9\,a}{128\,d}}\right )\lambda\,T
^{2}g^{2}\pi ^{-2} \nn \\
&&
+\left (-{\frac {9\,\ph^{2}}{4\,c}}-{\frac {3\,
\ph^{2}}{4\,d}}\right )\lambda^{2}T\pi ^{-1}g^{-1} \nn \\
&&
+\left (-{\frac {3
\,\ph^{2}c}{32\,d^{3}}}-{\frac {27\,\ph^{2}d}{32\,c^{3}}}+{\frac {3
\,\ph^{2}}{8\,d\left (d+c\right )}}+{\frac {9\,\ph^{2}}{32\,cd}}-{
\frac {15\,\ph^{2}}{32\,d^{2}}}\right )\lambda^{3}T^{2}\pi ^{-2}g^{-2
} \nn \\
&&
+\left (-{\frac {c^{2}}{256\,b\left (d+c\right )}}+{\frac {3\,b}{32
\,a}}-{\frac {d^{2}}{256\,b\left (d+c\right )}}+{\frac {cd}{128\,b
\left (d+c\right )}} \right. \nn \\
&&
\left.
+{\frac {\ph^{2}}{128\,b\left (c+b\right )}}+{
\frac {9\,d}{512\,a}}-{\frac {3\,\ph^{2}b}{128\,a^{3}}}-{\frac {3\,
\ph^{4}}{2048\,a^{3}\left (c+a\right )}}-{\frac {9\,\ph^{2}d}{2048\,
a^{3}}} \right. \nn \\
&&
\left.
-{\frac {3\,\ph^{2}c}{2048\,a^{3}}}+{\frac {\ph^{2}d^{2}}{
1024\,b^{3}\left (d+c\right )}}-{\frac {\ph^{4}}{512\,b^{3}\left (c+b
\right )}}-{\frac {\ph^{2}dc}{512\,b^{3}\left (d+c\right )}} \right. \nn \\
&&
\left.
+{\frac {
3\,\ph^{2}}{512\,a\left (c+a\right )}}+{\frac {\ph^{2}c^{2}}{1024\,b
^{3}\left (d+c\right )}}+{\frac {3\,c}{512\,a}}\right )T^{2}g^{4}\pi ^
{-2} \nn \\
&&
+\left (-{\frac {3\,\ph^{2}}{64\,a}}-{\frac {3\,\ph^{2}}{32\,b}
}-{\frac {3\,b}{8}}-{\frac {3\,a}{16}}\right )Tg^{3}\pi ^{-1} \nn \\
&&
+\left (
-{\frac {81\,\ph^{4}}{32\,c^{4}}}-{\frac {3\,\ph^{4}}{8\,d^{3}\left
(d+c\right )}}-{\frac {27\,\ph^{4}}{32\,c^{3}d}}\right )\lambda^{4}T^
{2}\pi ^{-2}g^{-4} \nn \\
&&
+{\frac {\left (-{\frac {3\,c}{4}}-{\frac {3\,d}{4}
}\right )\lambda\,Tg}{\pi }}.
\eea
According to (\ref{exactmass})
the masses determine the effective potential of order
$g^4$, $\lambda^2$ and $f_t^4$
\beq
V^{(3)}(\ph,T)=\int^\ph d\ph'\,{m_\chi^{(3)}}^2(\ph',T).
\eeq
Due to the complicated structure in $\ph$  of the integral, ($a,c$ and
$d$  are irrational functions of $\ph$) we have not evaluated
it analytically. Instead of it,  we have used
a numerical integration in order to determine the effective potential
at the critical temperature and to calculate the surface tension.

\vfill\eject

\appendix{\bf Appendix C \, Coset space volume}
\renewcommand{\theequation}{\mbox{C}.\arabic{equation}}
\setcounter{equation}{0}
\bigskip

Due to the global $SU(2)\times U(1)$ symmetry of the standard model,
which is spontaneously broken to the electromagnetic $U(1)$ subgroup,
the fluctuations around the saddle point (cf. eq. (\ref{saddle})) contain three
zero modes which have to be treated in the usual way
by the method of collective coordinates.

An infinitesimal transformation of the saddle point $\bar{\Phi}$,
which yields another saddle point with the same free energy, is
given by
\begin{equation} \label{coset}
d\Phi = dU \bar{\Phi}
= i\sum_{i=1}^3 d\omega_i T_i \bar{\Phi} \quad,
\end{equation}
where
\begin{equation} \label{matrix}
T_{1,2} = \frac{1}{2} \tau_{1,2}, \quad
T_3 = {1 \over 2} (\tau_3 - E) \quad;
\end{equation}
here $E$ denotes the identity matrix. Transformations generated by
$T_4 = {1\over 2}(\tau_3+E)$ leave the saddle point $\bar{\Phi}$
invariant. Comparing eq. (\ref{coset}) with
\begin{equation} \label{field}
d\Phi = \frac{1}{\sqrt{2}}
\left(
\begin{array}{c}
d\chi_1 + i d\chi_2 \\
i d\chi_3 \\
\end{array}
\right),
\end{equation}
yields the connection between the fields $d\chi_i$ and the group parameters
$d\omega_i$. The corresponding measure of the functional integral is
\begin{equation} \label{measure}
[d\chi]
= \prod^3_{i=1} \left(\frac{\beta}{2\pi}\right)^{1\over 2}dc_i,
\end{equation}
where $d\chi_i=\xi_i dc_i$, and $\xi_i$ is a normalized function, i.e.,
$\int d^3x\xi_i^2 = 1$. Eqs. (\ref{coset}), (\ref{field}) and
(\ref{measure}) yield
\begin{equation}
[d\chi] = -\frac{1}{4} \left(\frac{\beta}{2\pi}
\int d^3x \bar{\ph}^2\right)^{3\over 2}\prod^3_{i=1} d\omega_i.
\end{equation}

{}From eqs. (\ref{coset}) and (\ref{matrix}) one can easily calculate
the metric on the group manifold at the origin
$\omega=0$:
\beq
g_{ij} = - tr(U^{-1} \frac{\partial}{\partial \omega_i} U
              U^{-1} \frac{\partial}{\partial \omega_j} U)
       = c(i) \delta_{ij},
\eeq
where
\beq
c(1)=c(2)= {1\over 2},\quad c(3)=c(4)=1,
\eeq
\beq
\sqrt{detg(\omega=0)} = \frac{1}{2}\,. \label{norm}
\eeq
Globally, the symmetry group acting on the Higgs field is not
$SU(2)\times U(1)$, but rather $SO(3)\times Z_2\times SO(2)$,
where the elements of $Z_2$ are $E$ and $-E$, and $SO(2)$
corresponds to phase transformations with the phase varying from
$0$ to $\pi$.
For the normalization (\ref{norm}) the volume of this group is
(cf. ref. \cite{lue})
\begin{equation}
\int d\omega_1 \ldots d\omega_4 \sqrt{g} = 2 \pi^3.
\end{equation}
The Higgs vacuum expectation value breaks this symmetry to the
subgroup $U(1)$.
The corresponding volume of the coset space is
\begin{equation}
V(SO(3)\times Z_2\times SO(2) / U(1)) = \pi^2.
\end{equation}

For the integration over the zero modes one now obtains
\begin{equation}
\int [d\chi] = \frac{\pi^2}{2} \left( \frac{\beta}{2\pi}
\int d^3x \bar{\varphi}^2 \right)^{3\over 2}.
\end{equation}
The corresponding three eigenvalues in the symmetric phase are
$\sim \mu^2$, where $\mu=m_{\varphi}(0,T)$. Hence, the relative contribution
of the zero modes to the nucleation rate $\Gamma$ is
$\mu^3\int [d\chi]$.

\vfill\eject

\vfill\eject
\renewcommand{\thesection}{\ }
\section{Figure caption}

\begin{description}
\item[Figure \ref{oneloop}]: One-loop contributions
to the effective potential.

\item[Figure \ref{twoloop}]: Two-loop contributions
to the effective potential.

\item[Figure \ref{ring}]: Ring diagram contributions to the effective
potential with one-loop self-energy insertions.

\item[Figure \ref{selfenergy}]: One-loop self-energy contributions
to the gauge-boson.

\item[Figure \ref{oneloopfull}:] One-loop contributions
to the effective potential with full  propagators including
counterterms.

\item[Figure \ref{gap}:] The gap equations: All one-loop self-energy
corrections with full propagators.

\item[Figure \ref{gamma}:] The maximal value of the Higgs mass
for which the phase transition is first-order as function of $\g$.

\item[Figure \ref{xi}:]  The maximal Higgs mass as function
of the convergence parameter $\x$.
(The dependence on the top-quark mass
is smaller than the width of the line.)

\item[Figure \ref{effpot}:] The effective potential
to order $g^3$ and
$\la^{3/2}$ at the smallest temperature ($T^*$)
allowed for $\xi=2$.

\item[Figure \ref{surfacetension1}:] Surface tension as function
of the Higgs mass for different values of $\g$.

\item[Figure \ref{surfacetension2}:] Surface tension as function
of the Higgs mass with the $\lambda^{3/2}$-contributions (full line),
with $\lambda$-contributions only (long-dashed line)
and with contributions from the partial summation of $g^4$ and
$\lambda^2$ terms (short-dashed line).

\item[Figure \ref{ffluctpot}:] The potentials $U_\ph(r)$ and $U_\chi(r)$ for
the scalar fluctuations $\hat{\ph}$ and $\hat{\chi}$.

\item[Figure \ref{pend}:] The potential $\bar V$  for
three different Higgs masses at the corresponding nucleation
temperatures $T_e$.
The potential for $m_H=100\ GeV$ $(200\ GeV)$ has been multiplied
reduced by a factor
10 (factor 400).
\end{description}
\begin{list}{}{\usecounter{figure}}
\item \label{oneloop}
\item \label{twoloop}
\item \label{ring}
\item \label{selfenergy}
\item \label{oneloopfull}
\item \label{gap}
\item \label{gamma}
\item \label{xi}
\item \label{effpot}
\item \label{surfacetension1}
\item \label{surfacetension2}
\item \label{ffluctpot}
\item \label{pend}
\end{list}

\begin{thebibliography}{99}

\bibitem{wb1}
D. A. Kirzhnits and A. D. Linde,
Phys. Lett. B72 (1972) 471.

\bibitem{wb2}
S. Weinberg,
Phys. Rev. D9 (1974) 3357; \\
L. Dolan and R. Jackiw,
Phys. Rev. D9 (1974) 3320; \\
D. A. Kirzhnits and A. D. Linde,
Ann. Phys. 101 (1976) 195.

\bibitem{wb4}
V. A. Kuzmin, V. A. Rubakov and M. E. Shaposhnikov,
Phys. Lett. B155 (1985) 36.

\bibitem{wb5}
K. Enqvist et. al.,
Phys. Rev. D45 (1992) 3415.

\bibitem{wb6}
M. Dine et al.,
Phys. Rev. D46 (1992) 550.

\bibitem{wb7}
M. E. Carrington,
Phys. Rev. D45 (1992) 2933.

\bibitem{wb8}
C. G. Boyd, D. E. Brahm and S. D. H. Hsu,
preprint CALT-68-1795 (1992); \\
M. E. Shaposhnikov,
Phys. Lett. B277 (1992) 324;
Erratum-ibid. B282 (1992) 483.

\bibitem{u1}
W. Buchm\"uller, T. Helbig and D. Walliser, preprint DESY-92-151.

\bibitem{zwirner2}
J.R. Espinosa, M. Quir\'os and F. Zwirner, preprint CERN-TH.6577/92,
IEM-FT-58/92.

\bibitem{arnold}
P. Arnold, Phys. Rev. D46 (1992) 2628; \\
P. Arnold and O. Espinosa, preprint UW/PT-92-18 (1992).

\bibitem{parwani}
R. R. Parwani, Phys. Rev. D45 (1992) 4695.

\bibitem{jain}
V. Jain, preprint MPI-Ph/92-72 (1992); \\
V. Jain and A. Papadopoulos, preprint LBL-33067 (1992) ; \\
M. Carena and C. E. M. Wagner, preprint MPI-Ph/92-67 (1992).

\bibitem{wetterich}
N. Tetradis and C. Wetterich, preprint DESY 92-117 (1992); \\
M. Reuter and C. Wetterich, preprint HD-THEP-92-62 (1992).

\bibitem{patkos}
H. Meyer-Ortmanns and A. Patk\'os, Phys. Lett. B297 (1992) 321.

\bibitem{Connor}
D. O'Connor, C. R. Stephens and F. Freire,
preprint DIAS-STP-92-02 (1992).

\bibitem{latticesu2}
B. Bunk et al., Phys. Lett. B284 (1992) 371.

\bibitem{wb9a}
M. Gleiser, E. W. Kolb and R. Watkins,
Nucl. Phys. B364 (1991) 411; \\
N. Tetradis, preprint DESY 91-151,
Z. Phys. C, in press; \\
M. Gleiser and E. W. Kolb,
Phys. Rev. Lett. 69 (1992) 1304, \\
and preprint FERMILAB-Pub-92/222-A; \\
M. Gleiser and R. O. Ramos, Phys. Lett. B300 (1993) 271.

\bibitem{wb16}
J. Langer,
Ann. Phys. 41 (1967) 108; {\it ibid}. 54 (1969) 258.

\bibitem{wb12}
J. I. Kapusta,
Finite Temperature Field Theory, (Cambridge University Press,
Cambridge, 1989).

\bibitem{ewein}
E. J. Weinberg and A. Wu, Phys. Rev. D36 (1987) 2474.

\bibitem{frele}
B. A. Freedman and L. D. McLerran, Phys. Rev. D16 (1977) 1130.

\bibitem{linde}
A. D. Linde, Phys. Lett. 96B (1980) 289; \\
D. J. Gross, R. D. Pisarski and L. G. Yaffe, Rev. Mod. Phys. 53 (1981) 43.

\bibitem{kalas}
A. Billoire, G. Lazarides and Q. Shafi, Phys. Lett. B103 (1981);\\
J. E. Mandula and M. Ogilvie, Phys. Lett. B201 (1988) 117; \\
O. K. Kalashnikov, Phys. Lett. B279, (1992) 367.

\bibitem {wb18} %thinwall
A. D. Linde,
Nucl. Phys. B216 (1983) 421.

\bibitem{wb19}
C. Callan and S. Coleman,
Phys. Rev. D16 (1977) 1762.

\bibitem{wb20}
N. J. G\"unther, D. A. Nicole and D. J. Wallace,
J. Phys A: Math. Gen. 13 (1980) 1755.   %nur mit Langer67 zitieren.

\bibitem{buhel}
W. Buchm\"uller and T. Helbig, Int. J. Mod. Phys. C3 (1992) 799.

\bibitem{Csernai}
L. P. Csernai and J. I. Kapusta,
Phys. Rev. D46 (1992) 1379; \\
M. E. Carrington and J. I. Kapusta,
preprint TPI-MINN-92/55-T(1992).

\bibitem{wb21}
D. B\"odeker, W. Buchm\"uller, Z. Fodor and T. Helbig,
in preparation.

\bibitem{Pisarski}
E. Braaten and R. Pisarski, Nucl. Phys. B337 (1990) 569.

\bibitem{lue}
M. L\"uscher, unpublished notes; \\
R. Gilmore, Lie Groups, Lie Algebras, and Some of Their Applications
(Wiley \& Sons, New York, 1974).

\end{thebibliography}
\end{document}